\documentclass[11pt,preprint]{aastex}
\usepackage[usenames]{color}

\slugcomment{Draft Version}

\begin{document}


\title{High-Sensitivity 86\,GHz (3.5\,mm) VLBI Observations of M87:\\ Deep Imaging
of the Jet Base at a 10 Schwarzschild-Radius Resolution}

\author{
Kazuhiro Hada\altaffilmark{1,2}, 
Motoki Kino\altaffilmark{3}, 
Akihiro Doi\altaffilmark{4},
Hiroshi Nagai\altaffilmark{5}, 
Mareki Honma\altaffilmark{1,6},
Kazunori Akiyama\altaffilmark{1,7}, 
Fumie Tazaki\altaffilmark{1}, 
Rocco Lico\altaffilmark{2,8}, 
Marcello Giroletti\altaffilmark{2}, 
Gabriele Giovannini\altaffilmark{2,8}, 
Monica Orienti\altaffilmark{2}, and 
Yoshiaki Hagiwara\altaffilmark{1,9}
}

\begin{abstract}
We report on results from new high-sensitivity, high-resolution 86\,GHz (3.5
millimeter) observations of the jet base in the nearby radio galaxy M87, obtained
by the Very Long Baseline Array in conjunction with the Green Bank Telescope. The
resulting image has a dynamic range exceeding 1500 to 1, the highest ever achieved
for this jet at this frequency, resolving and imaging a detailed jet
formation/collimation structure down to $\sim$10 Schwarzschild radii ($R_{\rm
s}$). The obtained 86\,GHz image clearly confirms some important jet features
known at lower frequencies, i.e., a wide-opening angle jet base, a limb-brightened
intensity profile, a parabola-shape collimation profile and a counter jet. The
limb-brightened structure is already well developed at $<0.2$\,mas ($<28$\,$R_{\rm
s}$, projected) from the core, where the corresponding apparent opening angle
becomes as wide as $\sim$100$^{\circ}$. The subsequent jet collimation near the
black hole evolves in a complicated manner; there is a ``constricted'' structure
at tens $R_{\rm s}$ from the core, where the jet cross section is locally
shrinking. We suggest that an external pressure support from the inner part of
radiatively-inefficient accretion flow may be dynamically important in
shaping/confining the footprint of the magnetized jet.
We also present the first VLBI 86\,GHz polarimetric experiment for this source,
where a highly polarized ($\sim$20\%) feature is detected near the jet base,
indicating the presence of a well-ordered magnetic field. As a by-product, we
additionally report a 43/86\,GHz polarimetric result for our calibrator 3C\,273
suggesting an extreme rotation measure near the core.
\end{abstract}


\section{Introduction}
Accreting supermassive black holes at the center of active galaxies produce
powerful relativistic jets that are observed as a collimated beam of plasma, often
propagating beyond the host galaxies. Understanding the formation, collimation and
propagation of relativistic jets is a longstanding concern in
astrophysics~\citep{br1974, bz1977, bp1982, begelman1984}, and the recent
theoretical progress based on general relativistic magnetohydrodynamical
simulations has begun to elucidate the roles of the central black hole,
surrounding accretion flow and magnetic fields threading them, as well as the
mutual interactions among these components, in generating and collimating a
jet~\citep[e.g.,][]{mckinney2006, komissarov2007, mckinney2009, tchekhovskoy2011,
mckinney2012}. To test the implications from such theories and then to better
understand the jet formation, it is necessary to present a detailed observation
that can image the relevant scales.

\footnotetext[1]{Mizusawa VLBI Observatory, National Astronomical Observatory of Japan,
Osawa, Mitaka, Tokyo 181-8588, Japan; {\rm kazuhiro.hada@nao.ac.jp}}

\footnotetext[2]{INAF Istituto di Radioastronomia, via Gobetti 101, I-40129 Bologna,
Italy}

\footnotetext[3]{Korea Astronomy and Space Science Institute (KASI), 776 Daedeokdae-ro,
Yuseong-gu, Daejeon 305-348, Republic of Korea}

\footnotetext[4]{Institute of Space and Astronautical Science, Japan Aerospace
Exploration Agency, 3-1-1 Yoshinodai, Chuo, Sagamihara 252-5210, Japan}

\footnotetext[5]{National Astronomical Observatory of Japan, Osawa, Mitaka, Tokyo
181-8588, Japan}

\footnotetext[6]{Department of Astronomical Science, The Graduate University for
Advanced Studies (SOKENDAI), 2-21-1 Osawa, Mitaka, Tokyo 181-8588, Japan}

\footnotetext[7]{Department of Astronomy, Graduate School of Science, The University of
Tokyo, 7-3-1 Hongo, Bunkyo-ku, Tokyo 113-0033, Japan}

\footnotetext[8]{Dipartimento di Fisica e Astronomia, Universit\`a di Bologna, via
Ranzani 1, I-40127 Bologna, Italy}

\footnotetext[9]{Toyo University, 5-28-20 Hakusan, Bunkyo-ku, Tokyo 112-8606, Japan}

M87 is the first extragalactic jet discovered by Curtis nearly 100 years
ago~\citep{curtis1918}. This jet is exceptionally close to
us~\citep[$D=16.7$\,Mpc;][]{blakeslee2009}, and sufficiently bright across the
entire electromagnetic spectrum through radio to TeV $\gamma$-rays. These
observational advantages have allowed a broad range of studies associated with
relativistic-jet physics, including the large-scale jet
morphology~\cite[e.g.,][]{owen1989}, the nature of the optical jet emission and
shocks~\cite[e.g.,][]{biretta1999, perlman2001} and the origin of the high-energy
X-ray to $\gamma$-ray emission~\citep[e.g.,][]{harris2006, abramowski2012,
hada2014}. Moreover, optical measurements of the nuclear stellar dynamics suggest
the presence of a huge central black hole of $M_{\rm BH} = (6$--$6.6) \times
10^{9}M_{\odot}$~\citep{gebhardt2009, gebhardt2011}, although gas-dynamical
measurements derive a factor of two smaller $M_{\rm BH}$~\citep{ford1994,
harms1994, macchetto1997, walsh2013}. The combination of the proximity and the
large black hole yields a linear resolution down to 1\,milliarcsecond (mas) =
0.08\,pc = 140\,Schwarzschild radii ($R_{\rm s}$) (for $D=16.7$\,Mpc and $M_{\rm
BH} = 6 \times 10^{9}M_{\odot}$), which is typically 10 to 100 times finer than
those accessible in distant quasars or blazars. Therefore, M87 offers a privileged
opportunity for probing the launch/formation scales of a relativistic jet with
high-resolution Very-Long-Baseline-Interferometer (VLBI) observations.

With the recent advent of the global short-millimeter VLBI project so-called the
Event Horizon Telescope (EHT), observational studies of the M87 jet have become
possible at a spatial scale comparable to the event horizon. At 1.3\,mm (a
frequency of 230\,GHz) \citet{doeleman2012} resolved a jet base/launching
structure that has a size of 40\,$\mu$as, corresponding to $5.5$\,$R_{\rm s}$ (if
$M_{\rm BH} = 6.4\times 10^{9}M_{\odot}$ is adopted). More recently,
interferometric closure phases for the corresponding structure have been obtained
at this wavelength~\citep{akiyama2015}, allowing a comparison between the
observation and some horizon-scale theoretical models. Going to shorter
wavelengths is also beneficial with respect to the synchrotron opacity, since the
jet base becomes more transparent to the self-absorption
effect~\citep{konigl1981}. Indeed, the $\lambda^{+0.94}$ dependence of the M87
radio core position revealed by astrometric measurements suggests the 1.3\,mm core
to be located within a few $R_{\rm s}$ from the black
hole~\citep{hada2011}. Nevertheless, current VLBI experiments at such short
wavelengths are still technically challenging to synthesize interferometric images
due to the limited number of available antennas (thus only sparse $uv$-coverage)
as well as the severe atmospheric disturbance (thus shorter coherence time). This
prevents us from tracking the larger-scale propagation of the flow launched from
the central horizon-scale dimension, which is essential to fully understand the
subsequent acceleration and collimation of the jet.

So far, VLBI imaging studies of M87 have mostly been made at 7\,mm (43\,GHz),
1\,cm (22--24\,GHz), 2\,cm (15\,GHz) or longer. In these bands the M87 jet is
bright enough and a sufficient number of VLBI stations are available for allowing
an adequate $uv$-coverage. Previous high-quality VLBI images of the M87 inner jet
revealed some important features such as a wide-opening angle base, a
limb-brightened intensity profile and a counter jet~\citep{junor1999, ly2004,
ly2007, kovalev2007, hada2011}, as well as a detailed movie near the jet
base~\citep{walker2008}. More recently, the inner jet was found to sustain a
parabola-shape collimation profile over a wide range of distance from $\sim$100 to
$\sim$$10^{5}\,R_{\rm s}$ from the nucleus~\citep{asada2012, hada2013a,
nakamura2013}. However in these bands, it is impossible to achieve angular
resolution comparable to that of the short-millimeter VLBI unless relying on a
space-VLBI satellite~\citep{hirabayashi1998, dodson2006, kardashev2013}.  In
addition, the higher optical depth at such long wavelengths precludes us from
observing the close vicinity of the black hole.  As a result, there still remains
a large gap in our current understanding of this jet between the centimeter and
the short-millimeter VLBI scales.

In this context, an important ``bridge'' to connect this gap is observational
study at 3.5\,mm (86\,GHz). At present, 3.5\,mm is the shortest wavelength where
one can reliably obtain synthesized VLBI images, as represented by studies with
the Global Millimeter VLBI Array~\citep[GMVA; e.g.,][]{giroletti2008, lee2008,
agudo2007, molina2014, boccardi2015, hodgson2015, koyama2015}. The angular
resolution with 3.5\,mm VLBI is typically twice better than that at 7\,mm, and the
transparency to a jet base is also higher. On the other hand, compared to 1.3\,mm,
one can detect the extended (optically-thin) emission much further down the jet
due to the steep-spectral nature of the synchrotron radiation, which allows a
better monitoring of the larger-scale jet propagation. Therefore, the use of
3.5\,mm is currently an optimal choice in terms of angular resolution, opacity and
capability for imaging a jet. Nevertheless, there have been still less M87
observations in this band because 3.5\,mm VLBI is generally less sensitive than
that at 7\,mm due to the more rapid atmospheric fluctuations as well as the worse
effective aperture efficiency of telescopes. Thus the brightness of the M87 jet
base (typically hundreds mJy to $<$1\,Jy) may not be sufficient to detect
interferometric fringes at high signal-to-noise (SNR) ratios. Only a handful of
3.5\,mm VLBI images of M87 are published in the literature~\citep{lee2008,
rioja2011, nakamura2013}, and the obtained jet structure is not well characterized
since the dynamic ranges of their images are only a level of $\sim$100, although
somewhat better quality 3.5\,mm images showing a hint of limb-brightening are
presented in one of the early GMVA results~\citep{krichbaum2006}.

In this paper we report on results from new high-sensitivity, high-resolution
3.5\,mm VLBI observations of the M87 jet, obtained by the Very-Long-Baseline-Array
(VLBA) connected to the Green Bank Telescope (GBT). This is the first
GBT-incorporated 3.5\,mm VLBI observation for M87. The large collecting area of
GBT, together with the recent implementation of the National Radio Astronomy
Observatory's (NRAO) wideband recording system, has allowed us to obtain an
unprecedented-quality image of the M87 jet in this band, where the dynamic range
has improved by a factor of greater than 10 from the previous VLBA-alone 86\,GHz
images~\citep{rioja2011, nakamura2013}. In the next section we describe our
observations and data reduction. In Section~3 and 4, our new results are presented
and discussed, supplementarily using contemporaneous datasets at the lower
frequencies. In the final section we will summarize the paper. Throughout the
paper we adopt $D=16.7$\,Mpc and $M_{\rm BH}=6.0\times 10^{9}M_{\odot}$ for M87,
corresponding to 1\,mas = 0.08\,pc = 140\,$R_{\rm s}$. Spectral index $\alpha$ is
defined as $S_{\rm \nu}\propto \nu^{+\alpha}$. Also, any $\lambda$-related numbers
are described in frequency unit in the rest of the paper.

\begin{table}[ttt]
 \begin{minipage}[t]{1.0\textwidth}
  \centering 
  \caption{VLBA observations of M87} \medskip
    \scalebox{0.671}{
  \begin{tabular}{lccccccc}
    \hline
    \hline
    UT Date  & $\nu$ & Stations & $\Delta \nu$ & Beam size & $I_{\rm peak}$ &
			   $I_{\rm rms}$ & $I_{\rm peak}/I_{\rm rms}$ \\
             & (GHz) &  &  (MHz)  & (mas$\times$mas, deg.) & $\left(\frac{\rm mJy}{\rm beam}\right)$ &
			 $\left(\frac{\rm mJy}{\rm beam}\right)$ &  \\
          &   (a)   & (b) & (c)     & (d)     & (e)     & (f) & (g)   \\
    \hline
    2014 Feb 11............... & 86.266 & VLBA, GBT, $-$SC, $-$HN & 512 &$0.50\times 0.11, -12$ ($0.42\times 0.08, -13$) & 549 & 0.63 & 871\\
    2014 Feb 26............... & 86.266 & VLBA, GBT, $-$SC, $-$HN & 512 &$0.30\times 0.10, -10$ ($0.26\times 0.08, -11$) & 521 & 0.36 & 1447\\
\smallskip
    2014 Feb (11+26)$^{({\rm h})}$..  & 86.266 & VLBA, GBT, $-$SC, $-$HN & 512 & $0.38\times 0.11, -10$ ($0.28\times 0.08, -12$)& 547 & 0.29 & 1886\\
    2014 Mar  8................. & 43.230 & VLBA, $-$MK, $-$FD & 128 &$0.58\times 0.26, 25$ ($0.50\times 0.21, 27$) & 786& 0.85 & 924\\
                                 & 23.830 & VLBA, $-$MK, $-$FD & 128 &$0.94\times 0.48, 18$ ($0.84\times 0.39, 21$) & 880& 0.82 & 1073\\
    2014 Mar 26............... & 43.296 & VLBA & 256 & $0.43\times 0.21, 12$ ($0.35\times 0.17, 12$) & 795 & 0.50 & 1590\\
                               & 23.894 & VLBA & 256 & $0.70\times 0.37, 1$ ($0.60\times 0.30, -1$) & 926& 0.43 & 2153\\
    2014 May 8................. & 43.296 & VLBA & 256 & $0.40\times 0.21,-2$ ($0.31\times 0.16, -3$) & 735& 0.56 & 1312\\
                               & 23.894 & VLBA & 256 & $0.69\times 0.36, -6$ ($0.59\times 0.29, -7$) & 869 & 0.64 & 1358\\
    \hline
  \end{tabular} 
  }\medskip
  \end{minipage}
  \label{tab:img_prm} Notes: (a) central frequency; (b) participating
 stations. VLBA indicates all the ten VLBA stations. GBT, SC, HN, MK and FD are
 the Green Bank Telescope, Saint Croix, Hancock, Mauna Kea and Fort Davis,
 respectively. A minus sign before station name means the absence of that station;
 (c) total bandwidth; (d) synthesized beam with a naturally-weighting scheme. For
 reference, a beam size with a uniformlly-weighting scheme is also shown in
 bracket; (e) peak intensity of M87 images under naturally-weighting scheme; (f)
 off-source rms image noise level of M87 images under naturally-weighting scheme;
 (g) dynamic range calculated with $I_{\rm peak}$ and $I_{\rm rms}$; (h) combined
 visibility data over the two epochs.
\end{table}

\section{Observations and data reduction}
\subsection{86\,GHz data}
In February 2014 we observed M87 at 86\,GHz with VLBA in conjunction with
GBT. From VLBA eight out of the ten stations participated in this program since
the other two (i.e., Hancock and Saint Croix) do not have the 86\,GHz receiver. To
increase the overall sensitivity, we made an 8-hour-long quasi-full-track
observation twice on February 11 and 26. The observations were made in dual
(left/right-hand circular) polarization mode. The received signals were sampled
with 2-bit quantization and recorded at aggregate rate of 2\,Gbps (a total
bandwidth of 512\,MHz) using the digital-downconverter-4 (DDC-4) wideband
recording mode. The down-converted signals were divided into two 128\,MHz
sub-bands in each polarization respectively. As an overall system calibrator
(fringe check, bandpass, delay tracking, see below) of this program, we inserted
5-minute scans on the nearby bright source 3C\,273 (10\,degrees apart from M87 on
the sky) every 30 minutes. 3C\,273 was observed also for the purpose of antenna
pointing by adding another 6-minute-long scans every 30--60 minutes. The second
epoch has better weather conditions and system temperatures over the array. The
information on these data is summarized in Table~1.

The initial data calibration was performed with the Astronomical Image Processing
System (AIPS) developed at NRAO. We first corrected the visibility amplitude by
applying the measured system noise temperature and the elevation-gain curve of
each antenna. Atmospheric opacity corrections were made by solving for receiver
temperature and zenith opacity for each antenna. We then calibrated the amplitude
part of the bandpass characteristics for each station using the auto-correlation
spectra of 3C\,273.

The calibration of the visibility phase was made following the amplitude
calibration. To recover 86\,GHz fringes as much as possible, we performed the
phase calibration in the following way. We first corrected known phase variations
due to parallactic angle effects. Next, the instrumental phase and delay offsets
for each antenna were derived using a scan of 3C\,273, and subtracted from the
whole dataset assuming that they are constant with time (note that at 86\,GHz no
pulse-cal signals are available to calibrate instrumental effects). After that, we
ran a global fringe-fitting on 3C\,273 with its source structure model created by
the procedure described below (a point source model was assumed in the first round
where the source model was not available), and derived time evolutions of the
residual delay, rate and phase for each IF separetely. We detected fringes for
most of the scans at adequate SNRs. Since the derived residual delay component
(that mainly comes from unpredicted atmospheric fluctuations in the correlation
stage) slowly varies on the sky, we can use the 3C\,273's delay solutions as a
good first-order approximation for those of M87's residual delay. We thus
interpolated the derived 3C\,273's delay solutions to the scans of
M87. Implementing this process, now we can go to fringe-fitting on M87 with a
tight delay search window (as small as $\pm10$\,nsec), which is quite effective in
avoiding false signals. We performed a global fringe-fitting on M87 with a
solution interval of 15\,sec and an SNR threshold of 4.0. Since the phase and
delay offsets between IFs are already removed, we derived IF-averaged solutions
for M87 to increase SNR by a factor of 1.4. Similarly to 3C\,273, we ran two
cycles of fringe-fitting iteratively; the first round (where no source model
available) was performed with a point source model, and after creating a coarse
image, the second round was executed with the source model. This indeed slightly
increased the fringe detection rate. In every fringe-fitting the most sensitive
station GBT was used as the reference antenna.

\begin{figure}[ttt]
 \centering
 \includegraphics[angle=0,width=0.7\columnwidth]{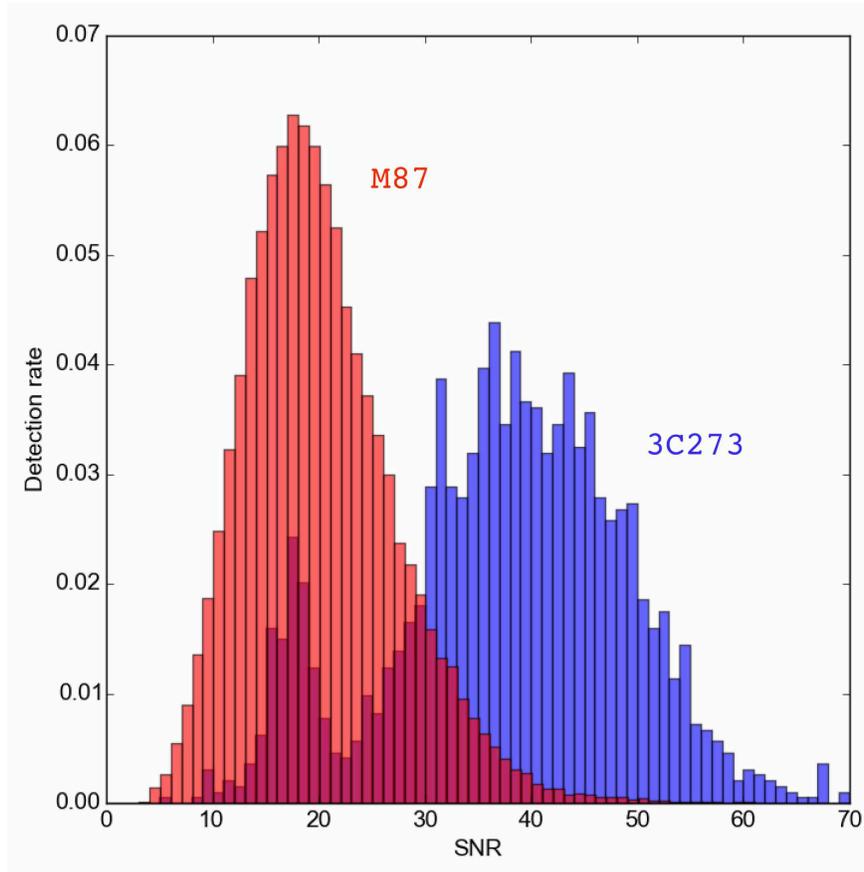}
 \caption{Normalized histograms of signal-to-noise ratio for the detected fringes
 (by global fringe fitting). The red histogram is for M87, while the blue one is
 for 3C\,273. A solution interval of 15\,sec is used here.} \label{fig:hist}
\end{figure}

Through this procedure, we recovered a number of fringes at sufficient SNRs. In
Figure~\ref{fig:hist} we show (normalized) histograms of SNR for the detected
fringes on M87 and 3C\,273. For 3C\,273 the median/mean SNRs were 38 and 38,
respectively. For M87 the median/mean SNRs were 19 and 20, respectively. Note that
the SNR histogram for 3C\,273 is relatively widely distributed. This is because
the source structure of 3C\,273 is highly complicated (see Appendix) and the
correlated flux density changes more drastically with baseline length, than in the
case of M87. 

After the visibility data became coherent with time and frequency, the data were
integrated over frequency (but the two IFs kept separated), and in time to 30
sec. These data were then used for creating images.

We performed our imaging process in the following way. We first exported the
averaged data to the Difmap software~\citep{shepherd1994}, and performed intensive
flagging of the visibility data. At 86\,GHz, antenna pointing is generally less
accurate than that at lower frequencies, and this causes an unwanted systematic
decrease of visibility amplitude in some scans. These scans then cause significant
sidelobes in the initial stage of imaging process, which prevents us from
reconstructing a reliable CLEAN/deconvolution model. Thus, in addition to obvious
outliers, we carefully flagged such bad scans in an antenna-based manner.

Following the initial exhaustive flagging, we then worked on iterative
CLEAN/self-calibration processing. We conducted this process using Difmap and AIPS
in a hybrid manner. We used Difmap only to perform CLEAN deconvolution (because
CLEAN with Difmap is faster and more intuitive than with AIPS). Then after
obtaining a reasonable set of CLEAN components, the model and the (uncalibrated)
visibility data were exported back to AIPS, and we performed self-calibration
using the AIPS task CALIB. This is advantageous for better calibration because the
self-calibration with CALIB can be handled more robustly and flexibly (e.g., in
terms of weighting for each antenna and SNR cutoff setting) than that in
Difmap. Self-calibration with CALIB was also necessary for proper polarization
analysis (see below) since CALIB can solve the complex gain terms for LL/RR
polarization separately, while Difmap cannot do that. Then the self-calibrated
visibility data were again exported back to Difmap to create an improved CLEAN
model. We repeated this CLEAN/self-calibration round until the reconstructed CLEAN
model no longer improved significantly. In the first several
CLEAN/self-calibration cycles, the phase-only self-calibration was done, and after
the phase part was well corrected, the visibility amplitude was also
self-calibrated iteratively with solution intervals starting from several hours
down to a few minutes. 

\begin{figure}[ttt]
 \centering \includegraphics[angle=0,width=0.7\columnwidth]{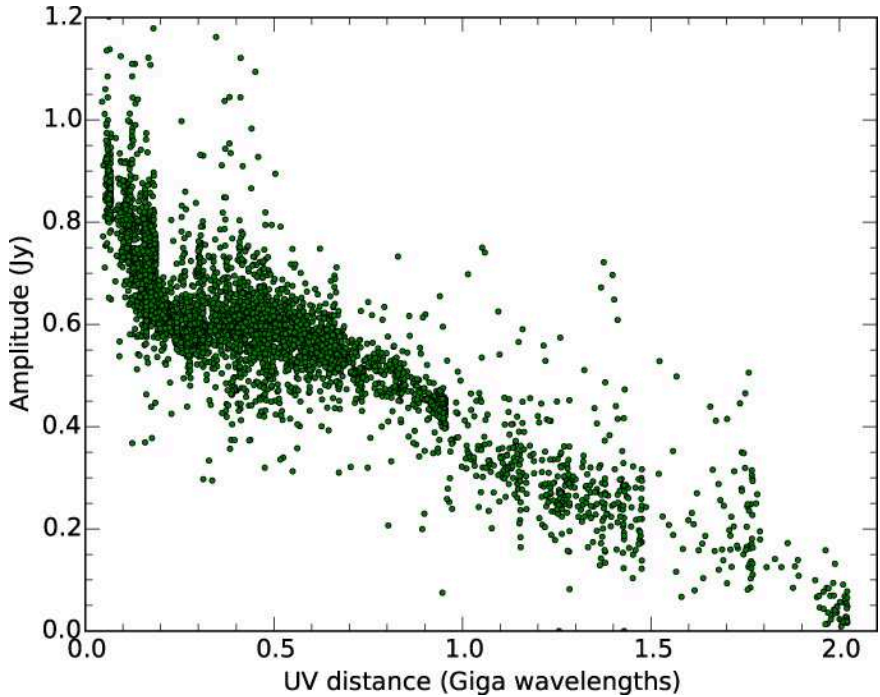}
 \caption{Visibility amplitude versus $uv$-distance plot of M87 for the VLBA+GBT
 86\,GHz observations. The visibility data shown are post-self-calibrated ones.}
 \label{fig:radplot}
\end{figure}

In Figure~\ref{fig:radplot} we show a resulting $uv$-distance plot of the
calibrated visibility amplitude of M87. Thanks to the excellent quality of the
dataset, one can see that the visibilities are homogeneously sampled over the
entire $uv$-distance and thus the overall trend is well-defined. One can see that
the whole visibility set primarily consists of two components. One is an extended
component which significantly contributes at baselines shorter than
$\lesssim$0.2\,G$\lambda$, while the other is a compact component that dominates
beyond 0.2\,G$\lambda$ to the longest 2\,G$\lambda$ baseline. For the compact
component, the correlated flux densities are monotonically decreasing with
increasing $uv$-distance, and the amplitude is $\lesssim$100\,mJy at the longest
baseline. Despite such low flux densities, the signals were robustly detected
since the longest baseline consists of GBT-MK pair.

\subsection{Polarization analysis at 86\,GHz}
Although our 86\,GHz observations were not ideally designed for accurate
polarimetric study, we can still attempt a polarization analysis by taking
advantage of the scans of 3C\,273. Following the parallactic angle corrections
described above, the cross-hand R-L phase and delay offsets were calibrated using
a scan of 3C\,273. The feed polarization leakage for each antenna was corrected by
using the LPCAL method in AIPS~\citep{leppanen1995} with a total intensity model
of 3C\,273. The derived leakage values were different with station, polarization,
IF and epoch, but we obtained a value of $9\pm2\,\%$ when averaged over the whole
array, which is consistent with the values derived in other VLBA 86\,GHz
polarimetric studies~\citep[e.g., $6\pm3$\%;][]{marti2012}. This is somewhat
larger than typical leakages obtained at lower frequencies, and roughly close to
the upper end where the linear approximation (``$D$-term'') scheme is
validated. GBT was one of the best stations where the leakages were as small as
$\sim$5\%, while a few of the other stations (particularly North Liberty and
Brewster) occasionally showed leakages up to $\sim$20\%. For each antenna we
confirmed an overall consistency (within a few percent) on the derived leakages
over the different IFs and epochs, indicating that the calibration is reasonably
valid. Following the prescription in \citet{roberts1994}, we estimate that the
corresponding error in fractional polarization in our images is $\sim$1\%. As a
check, we examined the minimum detectable polarization by using our 3C\,273
polarization map, and confirmed that a fractional polarization down to $\sim$0.7\%
was recovered at 3\,$\sigma$ in the core region (see Appendix). This is consistent
with the above calculation.

Regarding the electric vector polarization angle (EVPA), we cannot derive its
absolute value from our data alone, since we did not perform any additional
EVPA-calibration observation. However, we found a 43\,GHz VLBA observation of
3C\,273 that was carried out close in time with ours (on 2014 Feb 25) in the
Boston University blazar monitoring program. We used this 43\,GHz polarization
image as a reference of our EVPA correction (see Figure~\ref{fig:3c273} in
Appendix). Here we assume that the EVPA of the outermost polarized component (P3
at 1.5\,mas from the core) is stable with both time and frequency, and performed a
nominal correction of EVPA by matching the observed 86\,GHz EVPA of P3 to that of
the 43\,GHz one. Note that this assumption may not be correct, since the previous
concurrent 86/43\,GHz VLBA polarimetric study of this source suggests a large
rotation measure (RM) of $\sim$$2\times 10^{4}\,{\rm rad\,m^{-2}}$ for inner jet
components at $\sim$0.8\,mas from the core~\citep[see][]{attridge2005}. In fact, a
comparison of the present 86/43\,GHz polarization images similarly implies a
significant RM of the order of $\sim$$4\times 10^{3}\,{\rm rad\,m^{-2}}$ around P3
(see Appendix for more details). If this is the case, one would expect another
systematic rotation of the 86\,GHz EVPA about $\sim$8$^{\circ}$ with respect to
that at 43\,GHz (i.e., $\Delta \chi_{\rm RM}\sim 8^{\circ}$). As for the Boston
43\,GHz polarization image, we adopt its EVPA uncertainty to be $\Delta \chi_{\rm
43GHz} \sim 10^{\circ}$ based on \citet{jorstad2005}. Therefore, we estimate that
a potential total uncertainty of absolute EVPA in our 3C\,273's 86\,GHz images is
$\Delta \chi_{\rm 3C273} \sim \Delta \chi_{\rm 43GHz} + \Delta \chi_{\rm RM} \sim
\pm 18^{\circ}$. Regarding M87, its EVPA uncertainty in
86\,GHz images would be somewhat larger than this value, since the lower SNR of
polarization signals (SNR$\sim$4.5; see Section 3.4) from this source gives
another non-negligible thermal error term. This can be estimated by $\Delta
\chi_{\rm therm}({\rm radian}) \sim \sigma_{\rm p}/2P$ where $\sigma_{\rm p}$ and
$P$ are rms noise level and polarized intensity in polarization
map~\citep[e.g.,][]{roberts1994}. With SNR=$P/\sigma_{\rm p}$$\sim$4.5, we obtain
$\Delta \chi_{\rm therm,M87} \sim 6^{\circ}$. Assuming that $\Delta \chi_{\rm
3C273}$ and $\Delta \chi_{\rm therm,M87}$ are statistically independent, we
estimate a total error budget for M87 to be $\Delta \chi_{\rm M87} \sim \pm
20^{\circ}$.

\subsection{Lower frequency data}
As supplementary datasets, we additionally made VLBA-alone observations of M87 at
24 and 43\,GHz close in time with the 86\,GHz sessions. The observations were
carried out on March 8, 26 and May 8 2014, where both 24 and 43\,GHz were
quasi-simultaneously used by alternating each receiver quickly. On March 26 and
May 8, all the VLBA stations were present, while on March 8 the antennas at Mauna
Kea and Fort Davis were absent. We received only RR polarization signals with a
total bandwidth of 128\,MHz (on March 8) or 256\,MHz (on March 26 and May
8). Among these sessions, the data on March 26 were the best in overall quality,
while the data on March 8 were relatively poor. The initial data calibration
(apriori amplitude correction, fringe fitting and bandpass) was made in AIPS, and
the subsequent image reconstruction was performed in Difmap based on the usual
CLEAN/self-calibration procedure. The basic information of these data is also
tabulated in Table~1.

\section{Results}

\subsection{New 86\,GHz images}
\begin{figure*}[htbp]
 \centering
 \includegraphics[angle=0,width=1.0\textwidth]{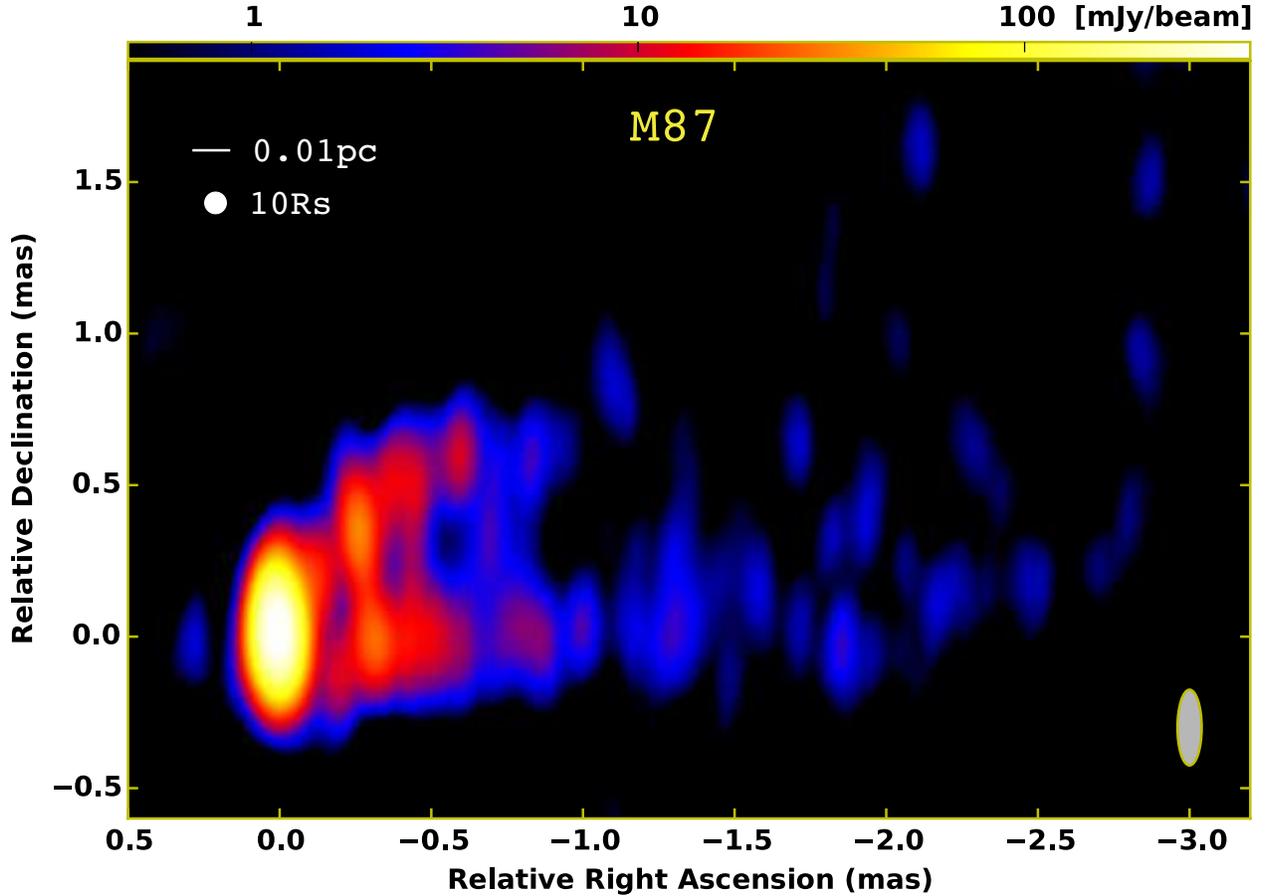}
 \caption{VLBA+GBT 86\,GHz false-color total intensity image of the M87 jet. The
 image is produced by combining the visibility data over the two epochs on 2014
 February 11 and 26. The restoring beam ($0.25\times0.08$\,mas in PA 0$^{\circ}$)
 is shown in the bottom-right corner of the image. The peak intensity is
 500\,mJy\,beam$^{-1}$ and the off-source rms noise level is
 0.28\,mJy\,beam$^{-1}$, where the resulting dynamic range is greater than 1500 to
 1. (A color version of this figure is available in the online journal.) }
 \label{fig:86GHzimage}
\end{figure*}

\begin{figure*}[htbp]
 \centering
 \includegraphics[angle=0,width=0.7\textwidth]{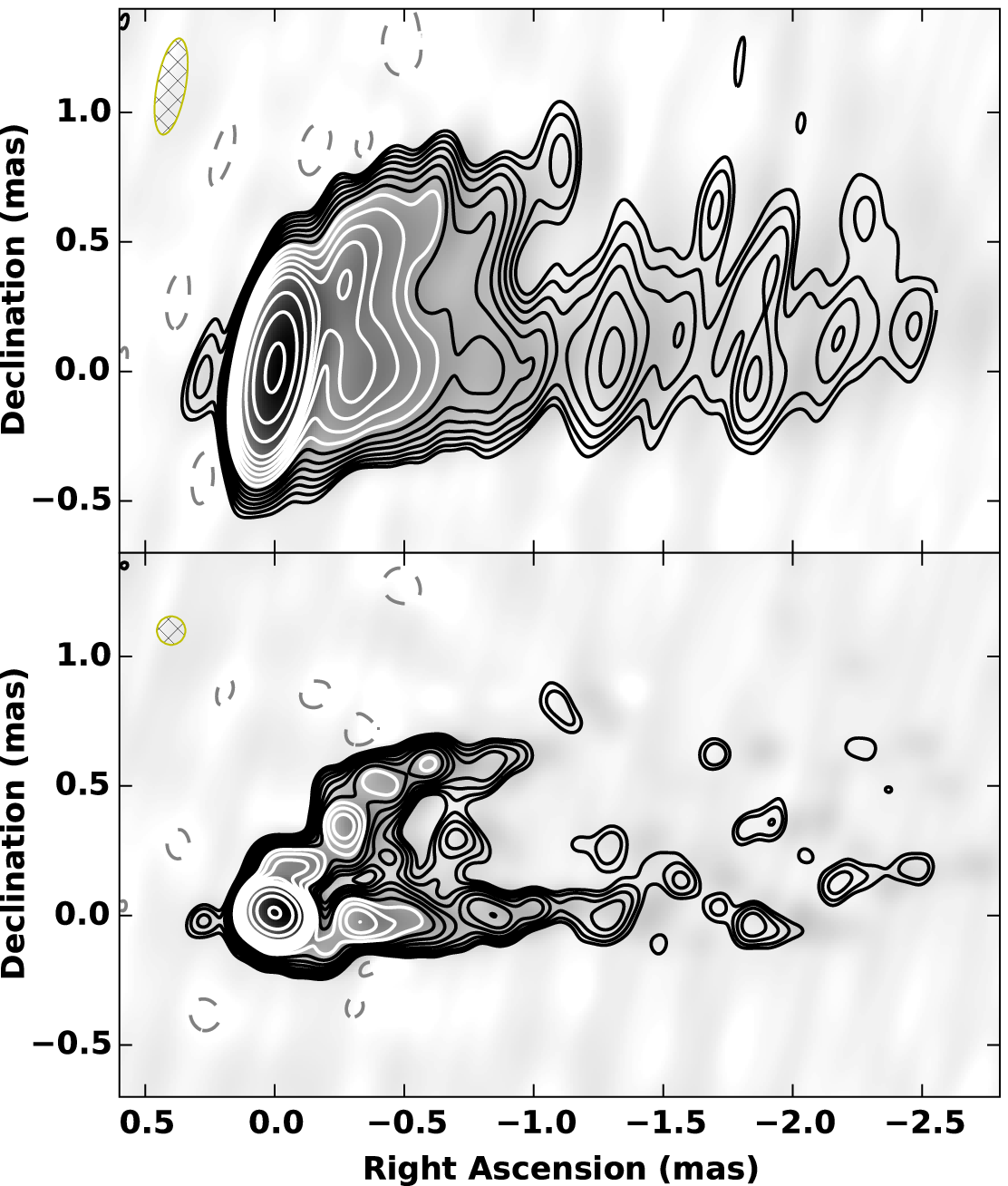}
 \caption{VLBA+GBT 86\,GHz total intensity contour images of the M87 jet. The
 upper panel indicates a naturally-weighted image with a synthesized beam of
 $0.37\times0.11$\,mas in PA $-10^{\circ}$. The lower panel shows a
 better-resolved image restored with a circular Gaussian beam whose FWHM is
 equivalent to that of the minor axis of the beam used in the top panel. In each
 image the restored beam is shown at the top-left corner of each panel. Contours
 in both images start from $-1$, 1, $2^{1/2}$, 2, 2$^{3/2}$, 4... times
 0.86\,mJy\,beam$^{-1}$.}  \label{fig:normal-super-compare}
\end{figure*}

In Figure\,\ref{fig:86GHzimage} we show a representative 86\,GHz image of the M87
jet obtained by our VLBA+GBT observations. For a better visualization, the image
is produced by combining the visibility data over the two epochs, and restored
with a convolving beam of 0.25\,mas $\times$ 0.08\,mas in a position angle (PA) of
0$^{\circ}$. A contour image with a naturally-weighting scheme is also displayed
in the top panel of Figure~\ref{fig:normal-super-compare}.

Thanks to the significant improvement of sensitivity, a detailed jet structure was
clearly imaged down to the weaker emission regions. The resulting image rms noise
of the combined image was $\sim$0.28\,mJy\,beam$^{-1}$. In this period the
extended jet was substantially bright down to $\sim$1\,mas from the core. The weak
emission was detected (particularly in the southern limb) down to $\sim$3\,mas
from the core at a level of 3\,$\sigma$, and another $\sim$1--2\,mas at
2\,$\sigma$ level. The peak surface brightness of the image was
500\,mJy\,beam$^{-1}$ at this resolution, corresponding to an image dynamic range
greater than 1500 to 1 (the detailed value slightly varies as a function of the
weighting scheme and convolving beam). This is the highest image dynamic range
obtained so far at 86\,GHz for this jet, and is quite comparable to typical
dynamic ranges in VLBA images at 43\,GHz~\citep[e.g.,][]{ly2007}. We describe a
comparison of our 86 and 43\,GHz images in the next subsection.

Regarding the individual epochs, the second epoch was better in overall image
quality than that at the first epoch. As listed in Table~1, the synthesized beam
for the first epoch is more elongated in the north-south direction than that for
the second epoch. This is mainly because the Brewster station, which constitutes
the longest baselines in the north-south direction, had the higher system noise
temperatures during the first session. Thus the less-weighted $uv$ dataset on this
station creates a slightly larger fringe spacing along the north-south direction.

Consistent with known lower-frequency images, most of the radio emission at
86\,GHz is concentrated on the compact radio core at the jet base. To quantify the
structure of the core region, we performed a single elliptical Gaussian
modelfitting to the calibrated visibility data with the Difmap task
\texttt{modelfit}. The derived model parameters are summarized in Table~2. As a
check we performed the same fitting to the three different datasets, i.e., the
Feb/11 data, the Feb/26 data and the combined one, but virtually the same result
was obtained. Additionally, we performed another elliptical Gaussian model fitting
on the image plane using the AIPS task JMFIT and examined the deconvolved result,
but this was also essentially the same within errors. The derived geometry of the
core is close to a circular shape with a diameter of $\sim$80\,$\mu$as, which is
just in between the sizes obtained at
230\,GHz~\citep[40\,$\mu$as;][]{doeleman2012} and
43\,GHz~\citep[110--130\,$\mu$as;][]{hada2013a}. A size of 80\,$\mu$as is
consistent with that obtained in our previous study based on an archival VLBA
86\,GHz dataset~\citep{hada2013a}, but the result presented here is much more
reliable. Adopting the parameters derived with the combined data, a brightness
temperature of the 86\,GHz core is estimated to be $T_{\rm B} = 1.8\times
10^{10}$\,K. Although the observed epochs are different, this value is quite
similar to $T_{\rm B}$ reported for the 230\,GHz core ((1.2--1.4)$\times
10^{10}\,{\rm K}$; Akiyama et al. 2015). No significant valiability was found in
$T_{\rm B}$ of the core between our two 86\,GHz sessions.

\begin{table}[ttt]
 \begin{minipage}[t]{1.0\columnwidth}
  \centering \caption{Modelfit parameters for 86\,GHz core}
    \begin{tabular*}{1.0\columnwidth}{@{\extracolsep{\fill}}ccccccc}
    \hline
    \hline
    Data  & $\theta_{\rm maj}$  & $\theta_{\rm min}$ & PA & $S_{\rm core}$  \\
          & ($\mu$as) & ($\mu$as) & (deg.) & (mJy)     \\
          &  (a)  &   (b)  & (c)  & (d)      \\
    \hline
    Feb 11 & $81\pm4$ & $62\pm10$ & $56\pm9$ & $669\pm67$   \\
    Feb 26 & $82\pm3$ & $81\pm7$ & $72\pm20$ & $652\pm65$  \\
    Feb (11+26)$^{({\rm e})}$ & $83\pm6$ & $76\pm4$ & $24\pm9$ & $672\pm67$   \\
    \hline
    \end{tabular*}
  \end{minipage}
  \label{tab:86Gcoremodel} Notes: (a) major axis size of the derived elliptical
  Gaussians; (b) minor axis size of the derived elliptical Gaussians; (c) position
  angles of the major axes of the Gaussian models; (d) total flux densities of the
  Gaussian models; (e) combined visibility data over the two epochs. For (a), (b)
  and (c), we estimate a practical uncertainty of each parameter by comparing the
  difference of the derived results between \texttt{modelfit} in Difmap and JMFIT
  in AIPS. For (d), we adopt 10\% uncertainty based on the absolute typical
  amplitude calibration accuracy.
\end{table}

Downstream of the core, we clearly identified a limb-brightened jet profile as
seen in Figure~\ref{fig:86GHzimage} and
Figure~\ref{fig:normal-super-compare}. While such a limb-brightened structure in
M87 is repeatedly confirmed in previous VLBA 43\,GHz/15\,GHz
images~\citep[e.g.,][]{junor1999, ly2007, kovalev2007, hada2011, hada2013a}, it was
not so clear in previous VLBA 86\,GHz images~\citep{rioja2011, nakamura2013},
although one of the early GMVA images presented by \citet{krichbaum2006} suggested
a hint of limb-brightening.

For a better description of the near-core structure, in the lower panel of
Figure~\ref{fig:normal-super-compare} we display the same image as in the upper
panel but restored with a circular Gaussian beam whose diameter is equal to the
minor axis of the synthesized beam in the upper panel. The jet launching
morphology within $\sim$0.5\,mas from the core appears to be complicated, but one
obvious feature found in this image is that the limb-brightened structure is
already well developed at $<0.2$\,mas (equivalent to $<28$\,$R_{\rm s}$,
projected) from the core with a large opening angle. We investigate the transverse
jet structure in more detail in Section~3.3.

At the eastern side of the core, we detected weak but significant emission (at a
level of 6\,$\sigma$ in the combined image) at $\sim$0.25\,mas from the core (see
the upper panel of Figure~\ref{fig:normal-super-compare}). This emission was also
detected in the second-epoch-only image, but was marginally seen in the
first-epoch-only image due partly to the lower sensitivity. The previous
high-dynamic-range imaging of this jet at 15 and 43\,GHz detected the counter
emission in more detail~\citep{kovalev2007,ly2007,acciari2009}, and they conclude
that the counter emission is a real counter jet. As described below, based on the
observed proper motion and consistent detection at 43\,GHz, we also conclude that
the counter feature detected in the present observations is a real counter jet
component emanated from the core.

\begin{figure}[ttt]
 \centering
 \includegraphics[angle=0,width=0.6\columnwidth]{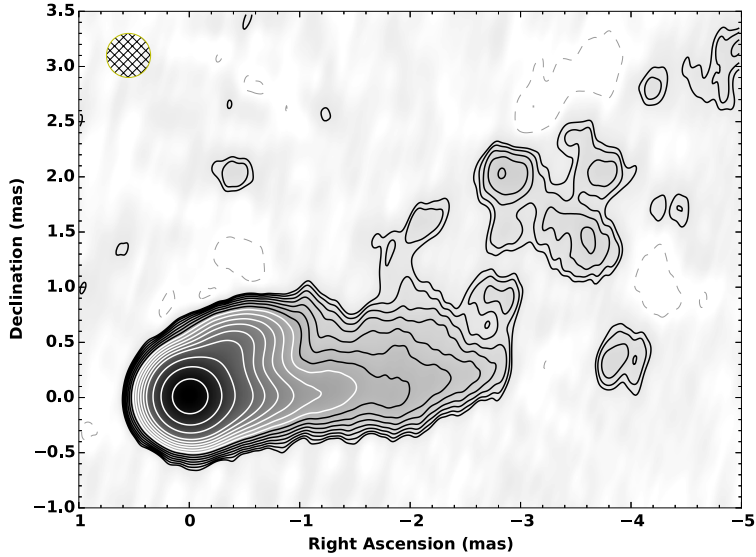}
 \caption{Tapered 86\,GHz image of the M87 jet. The image is convolved with a
 0.4\,mas diameter circular Gaussian beam. The contours are $-1$, 1, $2^{1/2}$, 2,
 2$^{3/2}$, 4... times 0.9\,mJy\,beam$^{-1}$.} \label{fig:m87taper}
\end{figure}

In Figure~\ref{fig:m87taper} we also show a tapered image restored with a
0.4-mas-diameter beam to emphasize the larger-scale emission more
noticeably. Consistent with known lower-frequency images, the 86\,GHz jet is
extending to the northwest direction on the large scale. In this period the
southern limb is brighter within $\sim$3\,mas from the core, then the northern
side rebrightens beyond that distance. Eventually the PA of the central jet axis
results in $\sim$293$^{\circ}$, matched with that in lower-frequency images.

\subsection{Comparison with lower frequency images}
Although our 86 and 43/24\,GHz observations were not simultaneous but 10 days to
10 weeks apart, it is still useful to compare them to examine any structural
consistency or variations. In fact, their comparable image dynamic ranges at a
level greater than 1000 to 1 allow a reliable image comparison between 86\,GHz and
the lower frequencies for the first time.

\begin{figure}[htbp]
 \centering
 \includegraphics[angle=0,width=0.55\columnwidth]{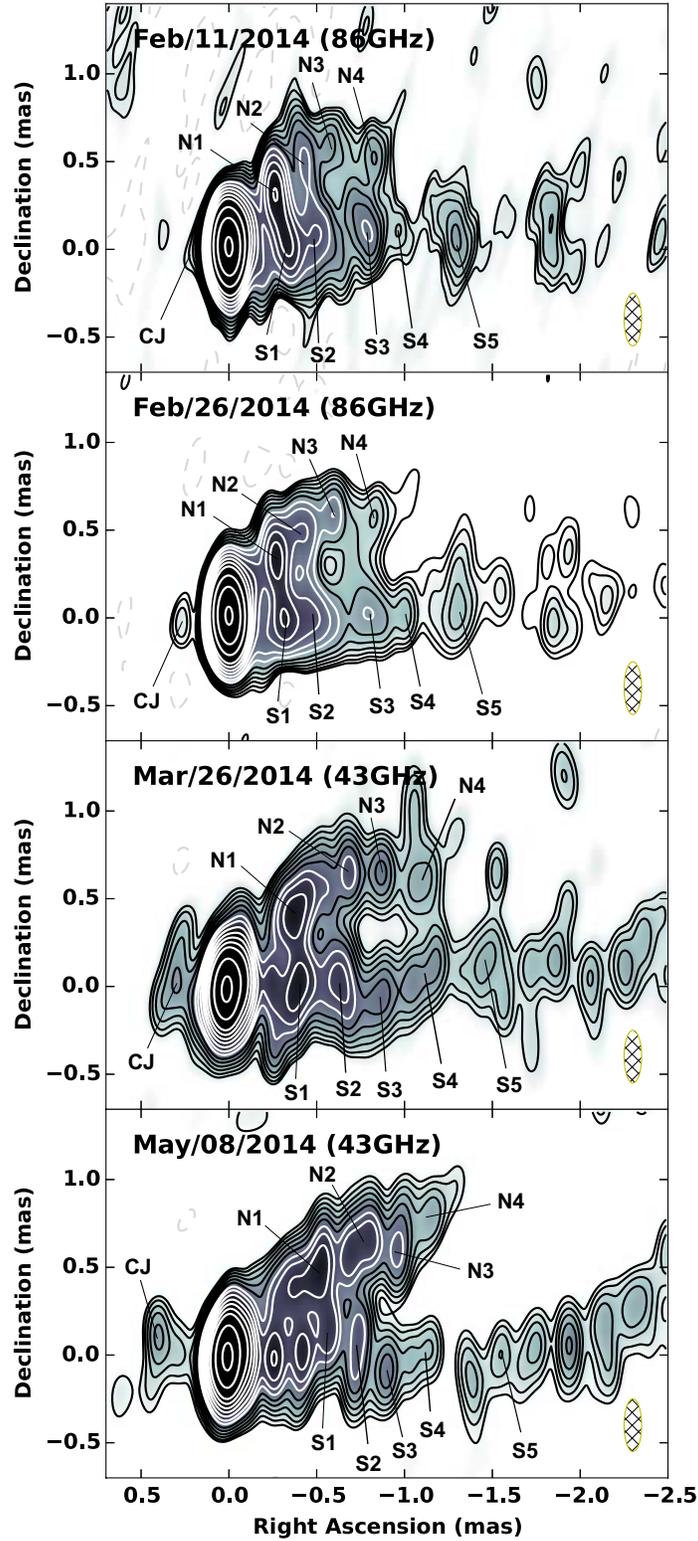}
 \caption{Multi-epoch images of the M87 jet. From the top, we show images observed
 on 2014 February 11 at 86\,GHz, 2014 February 26 at 86\,GHz, 2014 March 26 at
 43\,GHz and 2014 May 8 at 43\,GHz, respectively. All the images are convolved
 with a common beam of $0.30 \times 0.11$\,mas in PA 0$^{\circ}$ (shown at the
 bottm-right corner of each panel). Contours on each image are $-1$, 1, $2^{1/2}$,
 2, 2$^{3/2}$, 4... times 1.0\,mJy\,beam$^{-1}$ (upper two panels) and
 2.0\,mJy\,beam$^{-1}$ (lower two panels), respectively. The components with
 labels are the ones identified over the different epochs/frequencies
 consistently.}  \label{fig:m87wqcompare}
\end{figure}

\subsubsection{Jet morphology}

In Figure\,\ref{fig:m87wqcompare} we show a sequence of our 86 and 43\,GHz images,
where all the images are restored with the same convolving beam of $0.30 \times
0.11$\,mas in PA 0$^{\circ}$, which is approximately an intermediate resolution
between 86 and 43\,GHz. The overall jet shape and characteristic structure is in
good agreement with each other. It is known that the M87 jet is relatively smooth
and less knotty, but here we do see some noticeable features or patterns in the
jet when imaged at the high resolution. As shown in Figure~\ref{fig:m87wqcompare}
we identified several components in both 86 and 43\,GHz images consistently,
including a counter jet component (which is more prominent at 43\,GHz). These
features are marked as CJ (counter jet), N1--N4 (in the northern limb) and S1--S5
(in the southern limb), respectively. Since these components are typically
$\gtrsim$0.2\,mas/$\gtrsim$0.4\,mas apart from each other in east-west/north-south
directions, we could identify them separately regardless of the shape of applied
convolving beam between 86 and 43\,GHz. At 24\,GHz, its larger beam size led to
the mix of adjacent components, precluding us from identifying these inner
components separately. In our 24\,GHz images we instead identified another
component at $\sim$2\,mas downstream of the jet (in the northern limb), which is
marked as C24.

\subsubsection{Proper motions}
\begin{figure}[htbp]
 \centering \includegraphics[angle=0,width=0.7\columnwidth]{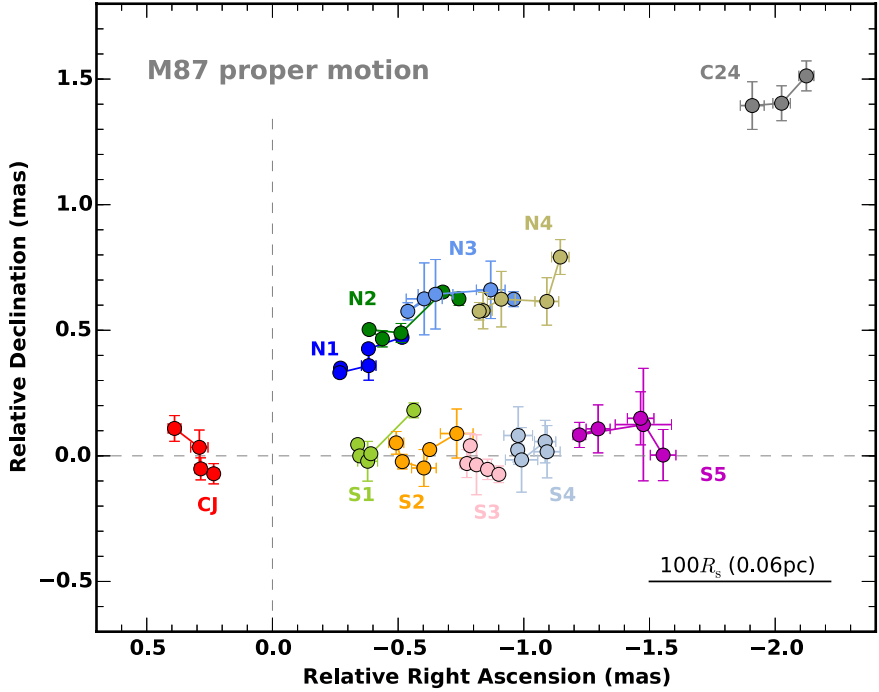}
 \caption{Observed sky positions of each component relative to the core. The
 components in the main jet are all moving toward the western direction, while the
 counter jet component CJ is moving to the opposite direction (i.e., to the
 northeast). } \label{fig:pos_comps}
\end{figure}

While these components were consistently identified over the observed period, they
were indeed gradually moving in the jet. In Figure\,\ref{fig:pos_comps} we show
the observed positions of these components with respect to the core over the
monitoring period. For each component the position was measured by fitting an
elliptical Gaussian model with the AIPS task JMFIT, and its position uncertainty
was estimated by the fitted size divided by the peak-to-noise
ratio~\citep{fomalont1999}. The measured proper motion results are summarized in
Table~3. Note that there may be an absolute position offset between the 86/43\,GHz
cores due to core-shift. However, such a shift is expected to be only
$\sim$20\,$\mu$as between 86/43\,GHz \citep[assuming the $\nu^{-0.94}$ dependence
determined at low frequencies;][]{hada2011}, which is substantially small compared
to the proper motions observed in the present study.

In the main jet a mean speed of the observed components is $\beta_{\rm app}=0.32$,
where most of them are moving at a similar speed in the range of 0.3--0.5. On the
other hand, the counter jet CJ is moving in the opposite (to the northeast)
direction at a slightly slower speed of $\sim$$0.17$\,$c$. These values observed
both in the jet and counter jet are similar to those suggested by \citet{ly2007},
although at that time the measurement was based on only one pair of VLBA 43\,GHz
observations separated by more than 6 months. Note that S3 and S4 appear to be
quite slow or quasi-stationary compared to the rest of the features in the main
jet. For these features we cannot completely rule out the possibility of component
misidentification among the different epochs. However, looking at the overall
evolution of the jet morphology within $\sim$1\,mas from the core, the northern
limb gets more elongated than the southern limb during our monitoring period (see
Figure~\ref{fig:m87wqcompare}). It therefore seems that an asymmetry exists in jet
apparent motions between the northern and the southern limbs, which is consistent
with the measured slow motions of S3 and S4.  The presence of quasi-stationary
components in the subpc jet of M87 is also suggested by \citet{kovalev2007} based
on a long-term VLBA 15\,GHz monitoring. We did not see clear signatures of the
faster ($\gtrsim$1$c$) motions as reported in \citet{walker2008} and
\citet{acciari2009}.

\begin{table}[htbp]
 \begin{minipage}[t]{1.0\columnwidth}
  \centering \caption{Component motion}
    \begin{tabular*}{1.0\columnwidth}{@{\extracolsep{\fill}}cccccc}
    \hline
    \hline
    Features  & $r$  & PA & $\mu$ &$\beta_{\rm app}$  \\
          & (mas) & (deg.) & (mas\,yr$^{-1}$) & ($v_{\rm app}/c$)   \\
      (a) &  (b)  &   (c)  & (d) & (e)     \\
    \hline
    CJ & $0.25\pm0.05$ & $106\pm8$ & $-0.63\pm0.29$ & $-0.17\pm0.07$   \\
    N1 & $0.44\pm0.01$ & $322\pm1$ & $1.22\pm0.08$ & $0.32\pm0.02$  \\
    N2 & $0.64\pm0.04$ & $317\pm2$ & $1.65\pm0.15$ & $0.44\pm0.04$   \\
    N3 & $0.86\pm0.16$ & $316\pm2$ & $1.45\pm0.16$ & $0.38\pm0.04$   \\
    N4 & $1.01\pm0.08$ & $304\pm4$ & $1.81\pm0.46$ & $0.48\pm0.12$   \\
    S1 & $0.34\pm0.02$ & $277\pm1$ & $1.00\pm0.02$ & $0.26\pm0.03$   \\
    S2 & $0.49\pm0.05$ & $275\pm2$ & $1.64\pm0.01$ & $0.43\pm0.03$   \\
    S3 & $0.78\pm0.05$ & $272\pm2$ & $0.31\pm0.18$ & $0.08\pm0.05$   \\
    S4 & $0.98\pm0.13$ & $274\pm2$ & $0.58\pm0.50$ & $0.15\pm0.13$   \\
    S5 & $1.29\pm0.11$ & $274\pm3$ & $1.25\pm0.61$ & $0.33\pm0.16$   \\
    C24 & $2.36\pm0.11$ & $306\pm3$ & $1.34\pm0.65$ & $0.35\pm0.17$   \\
    \hline
    \end{tabular*}
  \end{minipage}
  \label{tab:propermotion} Notes: (a) component name; (b)(c) radial angular
 distance and position angle of component from the core on 2014 February 11 (on
 March 8 for C24); (d) best-fit proper motion value; (e) corresponding apparent
 speed in the unit of the speed of light.
\end{table}

\subsubsection{Spectra}
Comparing the closest pair of the data on Feb/26 (at 86\,GHz) and
Mar/8 (at 43\,GHz), the observed peak brightness in each image (when measured with
a common $0.3\,{\rm mas} \times 0.1\,{\rm mas}$ beam as in
Figure~\ref{fig:m87wqcompare}) is 530\,mJy\,beam$^{-1}$ (at 86\,GHz on Feb/26) and
546\,mJy\,beam$^{-1}$ (at 43\,GHz on Mar/8), respectively. This results in a
non-simultaneous (but still close-in-time) 43/86\,GHz spectral index of the core
(at this resolution) being flat ($\alpha_{\rm c, 43/86}\sim$$-0.04$), indicating that
the M87 core is substantially self-absorbed up to 86\,GHz.

Considering the significant proper motions as well as the variable nature of the
individual components in the extended jet, it is difficult to obtain an accurate
spectral index distribution map with the present non-simultaneous 43/86\,GHz
images. Hence, here we only examined spatially-integrated spectra for the main and
counter jet in the following way. For the main jet, we compared the integrated
flux densities (sum of CLEAN components) between the two frequencies for an area
from 0.2 to 1.2\,mas along the jet, corresponding to the inclusion of N1--N4 and
S1--S4. We obtained an integrated spectral index of $\alpha_{\rm j, 43/86}=
-0.8\pm0.3$ for this region. For the same region, we also checked a spectral index
between 24 and 43\,GHz using our data, and obtained a value of $\alpha_{\rm j,
24/43}= -0.6\pm0.3$. These two values are consistent within the errors, although a
possible spectral steepening might exist at $>$43\,GHz due to the higher cooling
efficiency.

As for the counter jet, the observed spectrum between 43/86\,GHz seems to be
steeper than that of the main jet ($\alpha_{\rm cj, 43/86}\sim$$-1.8$). However,
the uncertainty is still large due to its weak nature.

\subsubsection{Jet to counter jet brightness ratio}

Given that the brightness profile of the main jet is very rich while the counter
jet is weak and less characterized, there is a large uncertainty in determining
the exact value of the jet-to-counter-jet brightness ratio (BR).

Here we consider the following possibilities to estimate BR: (1) CJ is the counter
part of S1: (2) CJ is the counter part of S2. The choice of (1) is because CJ and
S1 are symmetrically located with respect to the core, while the case (2) is also
possible if we additionally consider the relativistic ``arm-length ratio'' that
reflects a factor of $\sim$2 different apparent motions between the jet and the
counter jet.

In the case of (1), when compared using the total flux densities, BR results in
25.2 or 5.7 at 86 or 43\,GHz, respectively. In the case of (2), BR results in 22.0
or 4.4 at 86 or 43\,GHz, respectively. Alternatively, if we use the peak flux
densities, BR results in 25.3/4.7 at 86/43\,GHz for the case (1), while 21.4/5.6
at 86/43\,GHz for the case (2), respectively.

It is not clear yet whether the resulting difference in BR between 86 and 43\,GHz
is real or not. We regard that the measurements at 43\,GHz would be more secure
due to the higher SNR for the counter jet. Conservatively, here we conclude that
BR is between $\sim$5--25 within the central 1\,mas region.

\subsection{Transverse jet structure}
Resolving a jet in the direction transverse to the jet axis is important for
understanding the opening angle, collimation efficiency and possible velocity
gradient across the jet. To perform this, one needs a high quality image at a
sufficient angular resolution across the jet. The new 86\,GHz image presented here
allows us to analyze a detailed transverse structure of the jet launch region near
the black hole.

\begin{figure}[htbp]
 \centering \includegraphics[angle=0,width=0.9\columnwidth]{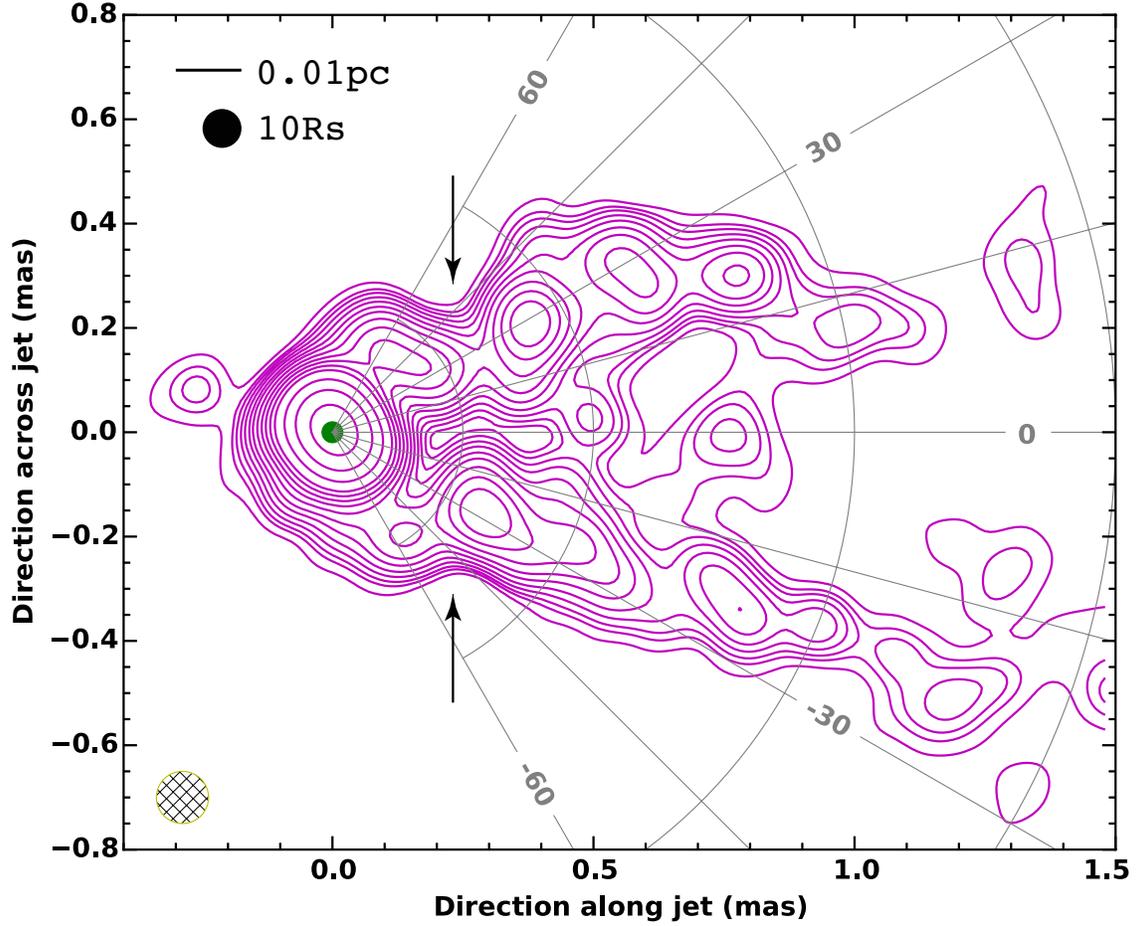}
 \caption{Close-up view of the innermost region of the M87 jet observed with
 VLBA+GBT at 86\,GHz. The image is restored with a 0.11\,mas circular Gaussian
 beam, which is equivalent to that of the minor axis of the synthesized beam. The
 image is rotated on the sky by $-23^{\circ}$ in order to align the jet central
 axis to the horizontal direction. For reference, on the origin of the coordinate
 we superpose a green-colored circle having a diameter of 40\,$\mu$as, which is
 equivalent to the size of the 230\,GHz core reported by \citet{doeleman2012}. The
 arrows point the place where the jet shape is locally shrinking.}
 \label{fig:openangleimage}
\end{figure}

In Figure~\ref{fig:openangleimage} we show a close-up view of the innermost region
of the M87 jet. For a better description across the jet, the image is restored
with a circular Gaussian beam whose FWHM is equivalent to that of the minor axis
of the synthesized beam. This is the same contour image as that in the lower panel
of Figure~\ref{fig:normal-super-compare}, but here the image is rotated on the sky
by $-23^{\circ}$ in order to align the jet central axis to the horizontal axis.

A strongly limb-brightened profile is evident beyond 0.5\,mas from the core, as
consistently seen in the 43\,GHz images. Within 0.5\,mas from the core, the
limb-brightening is continuously visible. It is notable that the limb-brightened
jet is already formed at 0.15\,mas distance from the core. Moreover, one more
remarkable feature near the core is that the limb-brightened jet is evolving in a
highly complicated manner; in particular, there is a ``constricted'' structure at
$\sim$0.2--0.3\,mas from the core, where the jet appears to be shrinking
locally. One might be concerned that this structure can be artificially produced
due to insufficient deconvolution accuracy, since the original synthesized beam of
our 86\,GHz array is elongated in the northeast-southwest direction, which
particularly could affect the near-core shape of the northern limb.  However, we
conclude that this is not an artifact. The root of the northern limb at
0.1--0.2\,mas from the core is relatively knotty, and we were required to put
relatively strong CLEAN components in our deconvolution process. Thus this feature
was distinguishable from the central core emission. As an independent test, we
also examined the fidelity of our super-resolution image at 86\,GHz using
sparse-modeling technique~\citep{honma2014}, and obtained a consistent structure
within 0.2\,mas from the core. Details about the application of this technique to
the present data will be described in a forthcoming paper. We additionally note
that the corresponding structure is marginally seen also in our 43\,GHz images (as
a ``neck'' of intensity between the core and N1/S1; see e.g., the middle panel of
Figure~\ref{fig:m87wqcompare}), suggesting that this structure is sustained at
least during our monitoring period (a few months).

\begin{figure}[htbp]
 \centering \includegraphics[angle=0,width=0.9\columnwidth]{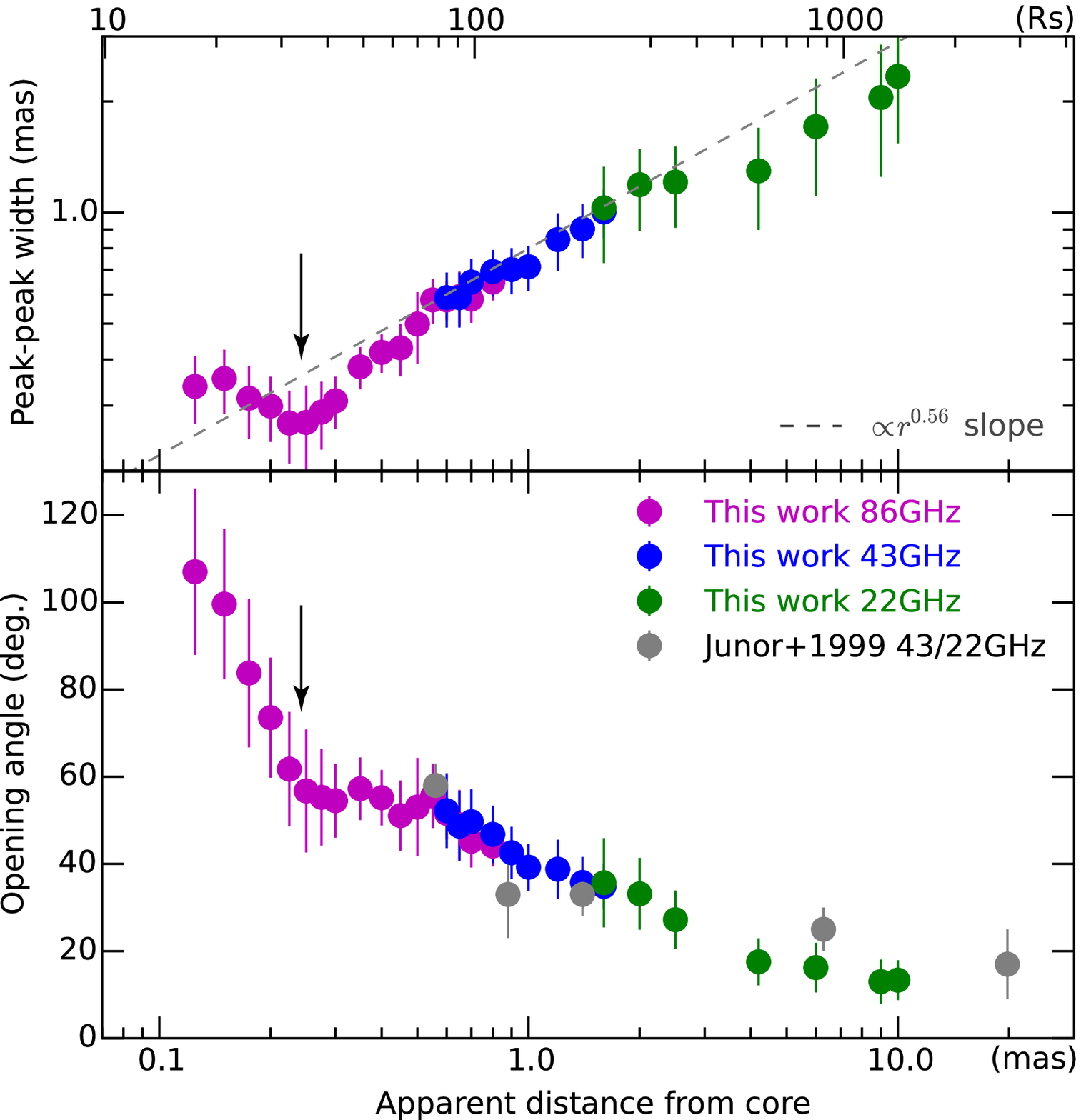}
 \caption{Radial profiles of the transverse structure of the M87 jet. The upper
 panel indicates a jet width profile $W_{pp}(r)$, which is obtained by measuring
 the peak-to-peak separation of the two limbs at each $r$ from the core. The grey
 dashed line represents $\propto r^{0.56}$ dependence, which was determined in our
 previous extensive jet width measurement between $r\sim$1\,mas and
 $r\sim$400\,mas from the core~\citep{hada2013a}. The dashed line plotted here is
 not a fitted one to the data but is arbitrarily placed just for reference. The
 lower panel shows the corresponding apparent opening angle profile $\phi_{pp}(r)
 \equiv 2\arctan(W_{pp}(r)/2r)$. The data points with grey color are previous
 results by \citet{junor1999} and \citet{biretta2002}. In both panels the data
 with magenta color, blue color and green color refer to 86, 43 and 24\,GHz,
 respectively. The arrow in each panel corresponds to the location of the arrow in
 Figure~\ref{fig:openangleimage}.} \label{fig:jetwidth}
\end{figure}

Using Figure~\ref{fig:openangleimage}, we analyzed the radial evolution of the
transverse jet structure in detail. The results are presented in
Figure~\ref{fig:jetwidth}, where we supplemented lower-frequency measurements at
43 and 24\,GHz further down the jet. The upper panel indicates the observed jet
width profile $W(r)$, while the lower panel plots the corresponding (apparent)
opening angle profile $\phi(r) \equiv 2\arctan(W(r)/2r)$ assuming the size of the
jet origin being infinitesimally small. As for $W(r)$ here we defined it by the
peak-to-peak separation of the two limbs perpendicular to the jet axis (thus for
clarity, we redenote the present measurements as $W_{pp}(r)$ and $\phi_{pp}(r)
\equiv 2\arctan(W_{pp}(r)/2r)$) rather than the usual outer-edge-to-outer-edge
separation that would be more appropriate for expressing an ``entire jet
width''. The reason why we use $W_{pp}$ here is because measurements of
peak-to-peak separation are less affected by the applied convolving beam, while
the measurements based on the outer edges are more sensitive to the beam
size. This is particularly relevant to the near-core region where the jet cross
section is comparable to or even smaller than that of the synthesized beam. Our
present purpose is not to determine an absolute width of the jet but rather to
measure a radial dependence of the jet evolution as close to the core as possible
by using a specific streamline in the jet. In this respect, $W_{pp}$ is a proper
way which permits to use a super-resolution image and thus to quantify the
streamline closer to the core. We note that the radial dependence of $W_{pp}(r)$
can be different from that of $W(r)$. This should be in fact an interesting topic
to be examined, but determining a detailed difference between them is beyond the
scope of our present work. Such an advanced analysis of the jet profile is indeed
possible by taking in advantage of the sparse modeling technique mentioned above,
and hence will be presented in the forthcoming paper.

The following interesting features are found in Figure~\ref{fig:jetwidth}. Beyond
0.5--0.6\,mas from the core, the measured jet shape is well described by a
parabolic collimation profile. There are several previous measurements for jet
width and opening angle on this scale. For the jet width, \citet{asada2012} and
\citet{hada2013a} determined a radial profile to be $W(r)\propto r^{0.56\pm0.03}$
over the distance from $\sim$1\,mas to $\sim$500\,mas from the core, and the
present $W_{pp}(r)$ beyond 0.5\,mas is in good agreement with this dependence. For
the opening angle, it was previously measured by \citet{junor1999} and
\citet{biretta2002}, and their finding of $\phi\sim 60^{\circ}$ opening angle at
0.5\,mas from the core is consistently seen in our result at the same
distance\footnote{In \citet{junor1999} and \citet{biretta2002} the opening angle
is defined with respect to the full-width-quarter-maximum (FWQM) on jet intensity
slice profiles. This gives a slightly larger opening angle than that in our method
at the same $r$.}. On the other hand, closer to the core where our 86\,GHz image
can access, the jet profile becomes more complicated. From the distance of
0.6\,mas to 0.3--0.2\,mas where the constriction exists, the observed opening
angle remains roughly constant at $\phi_{pp}(r)\sim 60^{\circ}$, indicating the
jet having a conical geometry. Then, even closer to the core within
$\sim$0.25\,mas, the jet widens again by a factor of 1.5--2. As a result, the
apparent opening angle rapidly increases from $\sim$$60^{\circ}$ up to
$\gtrsim$$100^{\circ}$ in this region. This trend can be recognized in
Figure~\ref{fig:openangleimage} with the guide of polar coordinates. If the
opening angle is defined with respect to $W(r)$ instead of $W_{pp}(r)$, the value
should be even larger.

Note that the opening angle may become smaller if the jet launch point has a
finite size (cross section). However, all the previous EHT observations of M87 at
230\,GHz constrained a jet-launch size to be remarkably
small~\citep[40\,$\mu$as;][]{doeleman2012, akiyama2015}. For reference, we
superpose the corresponding model of the 230\,GHz core on the coordinate origin of
Figure~\ref{fig:openangleimage}. As seen from this map, the modification of the
opening angle is sufficiently small at our scale of interest, so our assumption
should be reasonablly valid. Therefore, this is the largest opening angle ever
observed in any astrophysical jets as well as in M87 itself.

\subsection{Polarimetry}
\begin{figure}[htbp]
 \centering
 \includegraphics[angle=0,width=0.8\columnwidth]{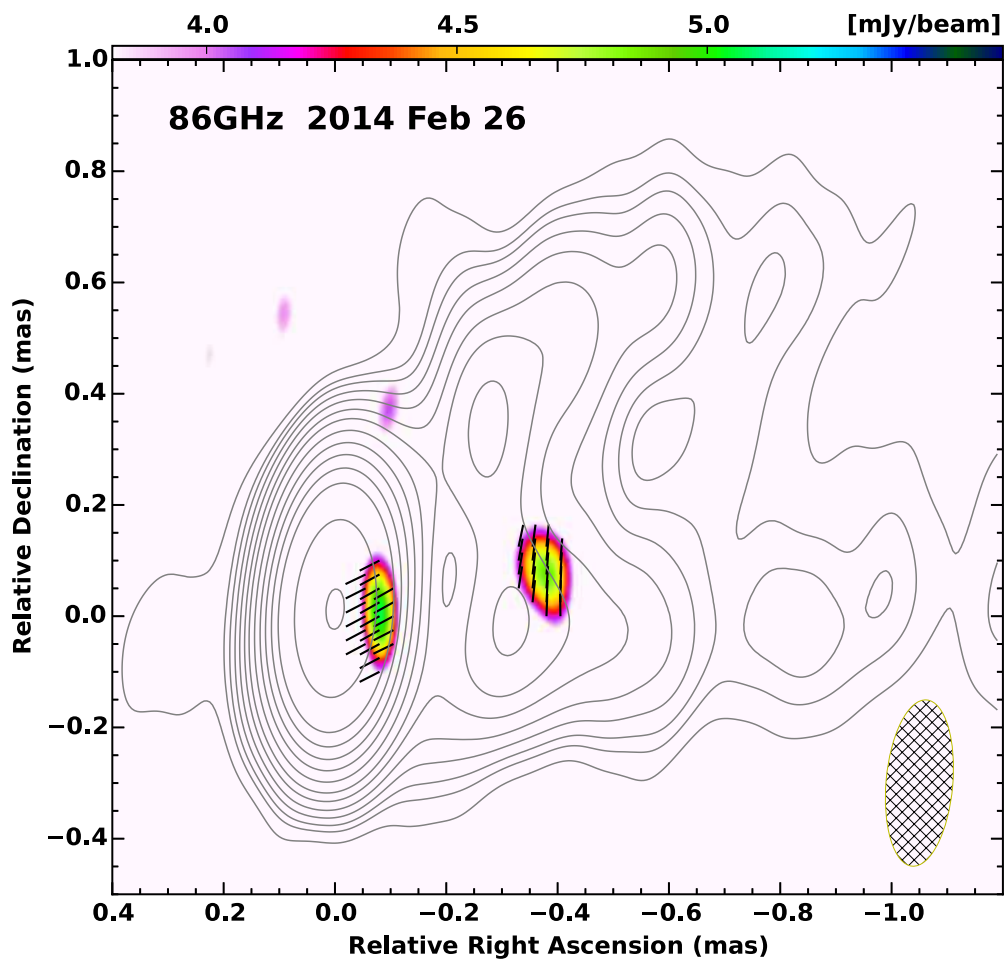}
 \caption{VLBA+GBT 86\,GHz polarimetric result for M87 on 2014 February 26. The
 color map, the vectors and the contours indicate the observed polarized
 intensity, the observed EVPA and the total intensity distribution,
 respectively. The convolving beam is shown at the bottom-right corner. The
 polarized intensity (color map) is displayed from 4\,$\sigma$ rms noise
 level. The contours start from 1, 2, $2^{3/2}$, 4... times 2.1\,${\rm
 mJy\,beam^{-1}}$.}  \label{fig:86Gpolmap}
\end{figure}

In Figure~\ref{fig:86Gpolmap} we show a result of our 86\,GHz polarimetry analysis
for the M87 jet. Here we display the data on February 26, since on this epoch the
data quality is better and also we can perform the more reliable EVPA correction
with the help of an external close-in-time VLBA observation of the calibrator
3C\,273 (see Appendix in more detail). The result on February 11 is essentially
consistent with Figure~\ref{fig:86Gpolmap}, although the SNR is lower than that on
February 26.

Given that the M87 jet is only weakly polarized (or highly depolarized) on
pc-to-subpc scales presumably due to a dense foreground Faraday
screen~\citep{zavala2002}, the present observation is still challenging to reveal
the whole polarimetric structure of this jet. Nevertheless, we indeed detected
some significant polarized emission in a few parts of the jet at SNR$\sim$4.5. One
polarized feature was detected at 0.1\,mas downstream of the core. This feature
has a polarized intensity of 5.0\,mJy\,beam$^{-1}$, and its fractional
polarization is a level of 3--4\%. On the other hand, we detected another
polarized feature at 0.4\,mas downstream in the jet. While this feature has a
similar polarized intensity (4.9\,mJy\,beam$^{-1}$) to that of the other one, the
observed fractional polarization becomes as high as 20\%. Since the previous VLBI
polarimetric observations of this jet (which are usually made at 15\,GHz or lower
frequencies) reported a fractional polarization up to 11.5\%~\citep{junor2001},
this is the highest fractional polarization ever reported on pc-to-subpc scales of
this jet. This polarized feature appears to be located at the boundary of S1 and
the central ``valley'' of emission (so-called the ``spine'' part of the jet).

Regarding EVPA, the near-core feature shows a mean EVPA direction along the jet,
while the near-S1 feature indicates a mean EVPA roughly perpendicular to the
jet. If this is the case, the corresponding magnetic-field-vector-polarization
angle (MVPA) of the near-core/near-S1 features are perpendicular/parallel to the
jet axis, respectively. However, we remind that there is still a large uncertainty
in our EVPA correction procedure ($\Delta \chi_{\rm M87}$$\sim$$\pm 20^{\circ}$;
see Section 2.2 or Appendix). Moreover, there might be an additional external EVPA
rotation if there is a significant amount of the foreground Faraday screen toward
M87 (e.g., $\Delta \chi_{\rm RM, M87} = {\rm RM}\,\lambda_{\rm
3.5mm}^2$$\sim$$7^{\circ}$ if RM$\sim$10$^4$\,rad\,m$^{-2}$). This does not permit
us to fix the intrinsic EVPA definitively.

\section{Discussion}
\subsection{Jet viewing angle and speed}
The viewing angle $\theta$ of the M87 jet has been discussed for a long time and
still remains as a puzzling issue for this jet. \citet{owen1989} suggest that the
jet is not too far out of the plane of the sky ($\theta \gtrsim 45^{\circ}$) based
on the patterns of the helically-wrapped filaments seen in the kpc-scale VLA
jet. On pc-to-subpc scales \citet{ly2007} suggest $\theta$$\sim$30--45$^{\circ}$
based on a proper motion and brightness ratio measurement with VLBA at 43\,GHz.
On the other hand, a strong constraint is obtained from optical observations of
the active knot HST-1 at $\sim$0.8$^{\prime\prime}$ from the nucleus, where
\citet{biretta1999} found a superluminal motion up to 6\,$c$, tightly requiring
$\theta$ to be smaller than 19$^{\circ}$ from our line of
sight. \citet{perlman2011} also suggest a similar range 
($\theta$$\sim$11--19$^{\circ}$) based on the optical polarization properties of
HST-1.

In the present study, we estimate $\theta$ near the jet base based on the measured
apparent motions in the jet and the counter jet, such that $\frac{\mu_{\rm
j}}{\mu_{\rm cj}} = \frac{1+\beta_{\rm int} \cos\theta}{1-\beta_{\rm int}
\cos\theta}$, in the assumption that the bidirectional jet is intrinsically
symmetric. If we compare CJ with S1 or S2 (as explained in Section~3.2.4), the
proper motion ratio results in $\mu_{\rm j}/\mu_{\rm cj} = $1.5--2.5. Searching
for a common area with $\mu_{\rm j} = \frac{\beta_{\rm
int}\sin\theta}{1-\beta_{\rm int}\cos\theta}$, we then obtain solutions of
$\theta$ and $\beta_{\rm int}$ to be $\theta = 29$--$45^{\circ}$ and $\beta_{\rm
int} = 0.29$--$0.50$, respectively. With these ranges of $\theta$ and $\beta_{\rm
int}$, we can also estimate an expected BR, and this results in a range of ${\rm
BR}=2.6$--$8.7$ (assuming a continuous jet model with $\alpha = -0.7$), which is
consistent with the observed BR in our 43\,GHz images.  Therefore, our estimate of
$\theta$ is rather similar to those suggested in \citet{ly2007}, whose measurement
was also made at a similar distance from the core. In contrast, the derived range
of $\theta$ is larger than that obtained from the optical HST-1 kinematics.

We do not rule out the possibility of the smaller $\theta$ as suggested from the
HST-1 observations, since there can be still an overlap $(\theta > 14^{\circ})$ if
we allow the maximum proper motion ratio $(\mu_{\rm j}/\mu_{\rm cj}=4.6)$ within
the 1\,$\sigma$ errors. Also, a recent VLBA 43\,GHz monitoring program of the
inner jet by Walker et al. suggests a fast apparent motion (in the main jet) of
$\sim$(1--2)\,$c$~\citep{walker2008}, which might favor a small viewing
angle. Thus the more accurate measurements of proper motion as well as brightness
ratio are important in future studies.

One issue, however, we should note is that the M87 jet is highly
limb-brightened. A commonly invoked explanation for limb-brightening structure in
relativistic jets is that there is a velocity gradient transverse to the jet, such
that the flow speed becomes faster toward the jet central
axis~\citep[e.g.,][]{ghisellini2005, nagai2014}. According to this idea, the
brighter part of the jet (i.e., the limb/sheath) has a larger Doppler factor
$\delta$ ($\delta \equiv [\Gamma (1-\beta \cos \theta)]^{-1}$ where $\Gamma$ is
the bulk Lorentz factor) to the observer, while the dim part of the jet (i.e., the
central spine) has a smaller $\delta$ due to the lower beaming relative to the
sheath. Since $\delta(\beta)$ reaches a maximum at $\beta = \cos\theta$,
if we consider the case of $\theta=10^{\circ}$--$20^{\circ}$, the faster $\beta$
yields the larger $\delta$ for most of $\beta$ unless $\beta_{\rm spine}$ is
unrealistically higher than $\beta_{\rm sheath}$. This results in $\delta_{\rm
spine} > \delta_{\rm sheath}$, indicating that the jet brightness profile would
lead to a \textit{ridge-brightened} structure. On the other hand, if we consider
the case of $\theta=29^{\circ}$--$45^{\circ}$, $\delta$ reaches a maximum at
$\beta$$\sim$0.70--0.87.  Then beyond this $\beta$, $\delta$ starts to decrease,
which in principle can reproduce a limb-brightened intensity profile.

Therefore, the observed limb-brightened structure of M87 may not be simply
explained by a transverse velocity gradient alone if
$\theta=10^{\circ}$--$20^{\circ}$, requiring some other process being at
work~\citep[e.g.,][]{lobanov2001, stawarz2002, gopal2007, zakamska2008,
clausen2011}.

An alternative hypothesis to accommodate the apparent discrepancy of $\theta$
between the inner jet and HST-1 is that the viewing angle of M87 is not constant
all the way down the jet. Our long-term VLBI monitoring of HST-1 has recently
revealed significant variations in the observed position angles (from
PA$\sim$270$^{\circ}$ to PA$\sim$310$^{\circ}$) of the substructures'
trajectories~\citep{giroletti2012, hada2015}\footnote{Changes in position angle of
parsec-scale superluminal components are also seen in another nearby radio galaxy
3C\,120~\citep{gomez2000, gomez2001,casadio2015}}. This implies a
deprojected (intrinsic) change in direction to be $\sim$10$^{\circ}$ (for a fixed
$\theta=15^{\circ}$). Thus, it would not be surprising if the $\theta$ of HST-1 is
also variable at this level, and the fastest $\sim$6\,$c$ speed could be seen when
its $\theta$ is maximally beamed to us. Such a local misalignment of $\theta$ from
the central jet axis can be realized if the HST-1 complex is traveling along a
3-dimensional helical trajectory with respect to the central jet axis.

From the point of view of the high-energy emission, the jet base of M87 is
proposed to be a likely site of the very-high-enery (VHE) $\gamma$-ray
production~\citep[e.g.,][]{aharonian2006, acciari2009, abramowski2012, hada2012,
hada2014}. The detection of VHE $\gamma$-ray emission usually prefers a small
viewing angle of the jet, which may also be opposed to the viewing angle derived
above. However, it would be interesting to note the jet base of M87 has a very
wide apparent opening angle up to $\phi_{\rm app}$$\sim$$100^{\circ}$
(Figure~\ref{fig:openangleimage} and Figure~\ref{fig:jetwidth}). If we consider
the case of $\theta=30^{\circ}$, the intrinsic opening angle is estimated
to be $\phi_{\rm int} \sim \phi_{\rm app} \times \sin\theta=50^{\circ}$. This
means that the near side of the sheath is almost pointing toward us at
$\theta_{\rm near}$$\sim$5$^{\circ}$ close to the jet base (assuming an
axially-symmetric jet). This value is quite similar to the typical viewing angle
in blazars. Therefore, the observed VHE emission from the jet base could be
associated with a locally beamed substructure in the near side of the
sheath~\citep[e.g.,][]{lenain2008, giannios2010}.

We also note that the bulk flow speed near the jet base may be somewhat faster
than what we measure in the component speeds. In fact, $\beta_{\rm int}$ derived
from the proper motion ratio $\mu_{\rm j}/\mu_{\rm cj}$ is applicable to both bulk
and pattern speeds, and one cannot distinguish between these two with such
measurement alone. A bulk flow speed is more directly related to the brightness
ratio BR. The expected BR of 8.7 described above (with $\beta_{\rm int}=0.5$ and
$\theta=29^{\circ}$) is actually only partially consistent with the present
observations (i.e., still inconsistent with the measured BR at 86\,GHz), and this
situation is essentially the same for the smaller angle of e.g.,
$\theta=15^{\circ}$ (${\rm BR}=11$). To reproduce the observed BR in our
43/86\,GHz images fully consistently, we suggest that the bulk flow speed needs to
be faster than $\sim$$0.6\,c$ at this location.

\subsection{Confinement of magnetized jet by hot accretion flow/corona?}
One of the most intriguing features found from our 86\,GHz observation is that the
initial jet formation structure evolves in a quite complicated manner; there are
multiple stages before the jet finally reaches the well-defined parabola in the
outer scale. The jet is formed with $\phi_{\rm app}\sim$100$^{\circ}$, then
rapidly collimated into $\phi_{\rm app}\sim$60$^{\circ}$ within $\sim$0.25\,mas (a
projected distance of $\sim$$35\,R_{\rm s}$) from the core, and subsequently
reaches the ``constricted'' point.  From there, the jet expands roughly conically at
$\phi_{\rm app}\sim$60$^{\circ}$ until $\sim$0.6\,mas ($\sim$84\,$R_{\rm s}$,
projected) down the jet, and finally enters the large-scale parabola collimation
zone. A possible structural change of the jet profile near the black hole is
originally 
suggested in \citet{hada2013a}. While this kind of feature might be self-formed
through some instabilities \citep[such as the sausage-pinch instability in a
magnetohydrodynamic flow; e.g.,][]{begelman1998}, this may also be a signature of
the interaction of the jet with the surrounding medium. In fact, there are growing
implications that a pressure support from an external medium is necessary to build
an efficient collimation of a jet~\citep[e.g.,][]{nakamura2006, komissarov2007,
komissarov2009}. Therefore, in what follows we discuss whether the formation of
the M87 jet on this scale can be subject to an external effect or not, based on a
simple comparison of the pressure balance between the jet ($p_{\rm jet}$) and the
external medium ($p_{\rm ext}$).

Here we consider the case that the internal pressure of the M87 jet is
approximated by the sum of the leptons ($p_{\pm}$) and the magnetic fields
($p_{B}$), and that the relative contribution from protons is
small~\citep{reynolds1996}. In \citet{kino2014}, we examined an allowed range of
the energy balance between electrons and magnetic fields at the base of this jet
based on the synchrotron self-absorption (SSA) theory, and showed that the radio
core at 43\,GHz can be highly magnetized or at most in roughly equipartition
($10^{-4} \lesssim U_{e}/U_{B} \lesssim 0.5$)\footnote{The equipartition range
cited here (that is originally from \citet{kino2014}) was derived, for simplicity,
in the assumption that the underlying magnetic-field configuration is isotropic
(the equation (11) in \citet{kino2014}). However, the same formula is also
applicable for the case of an ordered magnetic-field geometry just by multiplying
a small modification factor of $1/\sqrt{3}$ to the total field strength (see the
more detailed explanation in that paper).} unless the power index of the electron
energy distribution is too steep (i.e., unless $q>3$ where $n_{e}(E_{e}) \propto
E_{e}^{-q}$). The observed spectral index for the optically-thin part of the jet
in our 24/43/86 images ($\alpha_{\rm j}$$\sim$$-0.6$--$-0.8$ i.e.,
$q$$\sim$2.2--2.6 where $q \equiv -2\alpha+1$ in the present definition) satisfies
this condition. In this case, the total pressure of the jet is predominantly due
to the magnetic fields, or the particle pressure is at most of the same order of
magnitude of the magnetic one.  Therefore, we can reasonably adopt that $p_{\rm
jet} \sim p_{B} = B^2/8\pi$ at the 43\,GHz core. As for the 86\,GHz core, we can
similarly estimate its $B$ value based on the SSA formula~\citep[the equation (11)
in][]{kino2014} in combination with the modelfit parameters listed in Table~2, and
this results in $B_{\rm core, 86}$$\sim$8.3\,G. This value is just in between
$B_{\rm core, 43}$ and $B_{\rm core, 230}$ derived in \citet{kino2014, kino2015},
and a magnetically-dominated state can be consistently satisfied. Thus adopting
this $B_{\rm core, 86}$, we obtain $p_{B,\,\rm{core, 86}} \sim 2.7\,
\rm{dyn\,cm^{-2}}$. For the jet downstream of the core, $p_{B}$ depends on the
radial profile of the magnetic fields. In any case, to support/confine the jet on
these scales, the external medium needs to have a pressure that can be balanced
with the suggested level of of $p_{\rm jet}$.

According to the observed $\nu^{-0.94}$ frequency dependence of the core
shift~\citep{hada2011}, the radio core at 86\,GHz is estimated to be located at
$\sim$3\,$R_{\rm s}$ from the black hole on the plane of the sky. This indicates a
deprojected distance of the 86\,GHz core to be between $\sim$6\,$R_{\rm s}$ and
$\sim$12\,$R_{\rm s}$ for a range of $\theta=15^{\circ}$ or $\theta=30^{\circ}$,
respectively. On these scales a likely source of the external confinement medium
may be the inner part of accretion flow or associated coronal
region~\citep{mckinney2006, mckinney2007}. Since the accretion rate onto the M87
nucleus is significantly sub-Eddington~\citep{dimatteo2003}, the accretion mode of
the M87 black hole is thought to be an advection-dominated, hot accretion flow
state~\citep[ADAF; e.g.,][]{narayan1994}. As ADAF is geometrically thick and
roughly approximated by a spherically symmetric structure, such a configuration
may be suitable in shaping/confining the initial stage of a jet. While the
original ADAF assumed a radially-constant mass accretion rate ($\dot{M}$),
subsequent theoretical studies favorably suggest that $\dot{M}$ decreases with
decreasing radius due to convection~\citep{quataert2000} or
outflows~\citep{blandford1999}. In fact, the recent polarimetric study for the M87
nucleus at 230\,GHz has derived $\dot{M}$ at $r$$\sim$$21\,R_{\rm s}$ from the
black hole to be $\dot{M}< 9.2\times
10^{-4}$\,$M_{\odot}$\,yr$^{-1}$~\citep{kuo2014}, which is more than 100 times
smaller than that measured at the Bondi radius~\citep[$\dot{M}_{\rm
Bondi}\sim$0.1\,$M_{\odot}$\,yr$^{-1}$ at $r_{\rm Bondi}\sim$230\,pc$\sim$$4\times
10^5\,R_{\rm s}$;][]{dimatteo2003}. For such modified ADAF flows, \citet{yuan2012}
and \citet{yuan2014} present an updated set of self-similar solutions by taking
into account radially-variable $\dot{M}$ (i.e., $\dot{M}(r) \propto r^{s}$ where
$s\sim$0.5--1). Among these solutions, the pressure profile is described as
$p_{\rm ADAF}(r) \approx 1.7\times 10^{16} \alpha_{\rm visc}^{-1}m_{\rm
BH}^{-1}\dot{m}_{\rm BH}r^{-5/2+s}$\, dyn\,cm$^{-2}$, where $\alpha_{\rm visc}$,
$m_{\rm BH}$ and $\dot{m}_{\rm BH}$ are the dimensionless viscosity parameter,
$M_{\rm BH}/M_{\odot}$ and $\dot{M}_{\rm BH}/\dot{M}_{\rm Edd}$ ($\dot{M}_{\rm
Edd}\equiv L_{\rm Edd}/0.1c^2$), respectively\footnote{The self-similar solutions
presented in \citet{yuan2012}, \citet{yuan2014} and also in \citet{narayan1994}
(for the original ADAF) are obtained using a height-integrated system of
equations. Thus ``$r$'' in this case corresponds to the cylindrical radius, not
the spherical radius from the central black hole.  However, the effect of the
vertical integration is examined in detail by \citet{narayan1995a}, and they
proved that the height-integrated solutions are quite accurate approximations
(within $\sim$10\% for the pressure) of the exact (sperically-averaged) solutions
in the limit of advection-dominated state. Hence, we can reasonably treat that $r$
in the height-integrated solutions corresponds to the radial distance from the
black hole.}. For $\dot{M}_{\rm BH}$ onto the black hole, here we assume
$\dot{M}_{\rm BH}(r=R_{\rm s})\sim$$10^{-4}M_{\odot}{\rm yr}^{-1}$ together with
$s=0.6$. These values are selected so that the extrapolated values of $\dot{M}(r)$
at $r=r_{\rm Bondi}$ and $r=21\,R_{\rm s}$ consistently satisfy the above
observations, and thus would be a reasonable combination of $\dot{M}_{\rm BH}$ and
$s$. With these values, we finally obtain $p_{\rm ADAF}(r) \sim 2.2\alpha_{\rm
visc}^{-1}r^{-1.9}$\, dyn\,cm$^{-2}$. Therefore, if we assume $\alpha_{\rm visc}$
to be of the order of $10^{-2}$~\citep[e.g.,][]{yuan2012}, the ADAF pressure
results in $p_{\rm ADAF}\sim$ 2.7\,dyn\,cm$^{-2}$ at $r=10\,R_{\rm
s}$. Interestingly, this is quite comparable to $p_{\rm jet}$ estimated at the
86\,GHz core. Hence, we suggest that the pressure support from the inner part of
the hot accretion flow may be dynamically important in shaping and confining the
launch stage of the M87 jet.

Given that the external pressure contribution is significant at the jet base, the
observed constricted structure at a projected distance of $\sim$35\,$R_{\rm s}$
from the core may reflect some important physical signature resulting from the
jet-surrounding interaction. One possibility is that this feature marks a
reconfinement node of the flow~\citep[e.g.,][]{daly1988, gomez1997,
komissarov1997, kohler2012a, matsumoto2012, mizuno2015}. Such a flow reconfinement
can be realized if the radial profile of the external pressure downstream of the
core decreases more slowly than that of the jet pressure. Alternatively, this
constriction might be a signature of the sudden ``breakout'' from the central
dense confining medium (presumably ADAF/hot corona), analogous to a jet in
Gamma-Ray Bursts~\citep[e.g.,][]{morsony2007}. If the breakout is the case, one
may constrain a scale height (thickness) of the central confining medium to be of
the order of $H$$\sim$$2\times 35\times(\sin\theta)^{-1}=140\,R_{\rm s}$ (for
$\theta=30^{\circ}$).

We note that the above discussion still leaves a considerable uncertainty in each
parameter space and the actual force balance at the jet boundary may be more
complicated (e.g., if the jet ram pressure at the boundary is significant or if
there exists some instability in the accretion flow). In particular, the deep
86\,GHz images obtained here evoke the following simple question: \textit{why is
the suggested hot accretion flow still not detected in emission, despite the fact
that an accretion scale well below 100\,$R_{\rm s}$ near the black hole is already
imaged?} This issue should be explored by future higher-sensitivity imaging
observations~\citep[e.g., including the phased-up ALMA;][]{fish2013}. Also, the
lack of additional 86\,GHz images at different epochs with a similar sensitivity
does not permit us to conclude whether the observed constricted feature is a
persistent structure or just a temporal one (although we note that the subsequent
43\,GHz images in Figure~\ref{fig:m87wqcompare} still show a hint of the
corresponding feature at the same location). In any case, our simple
order-of-magnitude estimate discussed here implies a non-negligible contribution
of the external medium to the initial evolution of the M87 jet. Further VLBI
monitoring for the jet base at 86\,GHz will enable us to address this issue more
definitively.

\subsection{Implications for Faraday screen and magnetic field near the jet base}
Finally we briefly discuss the polarization structure of the M87 jet. While on kpc
scales polarimetric properties of this jet are intensively studied in radio and
optical~\citep[e.g.,][]{owen1989, perlman1999, chen2011, perlman2011}, on
pc-to-subpc scales the polarization structure is still highly uncertain. This is
because polarization signals from the M87 inner jet are generally quite low
presumably due to a dense foreground medium on these scales, as suggested by
\citet{zavala2002}.  They found an extreme RM distribution that varies from
$-4\times 10^3\,{\rm rad\,m^{-2}}$ to $>9 \times 10^3\,{\rm rad\,m^{-2}}$ in the
jet at 20\,mas (1.6\,pc, projected) from the core. Recently, \citet{kuo2014}
performed the first RM measurement toward the nucleus in the 230\,GHz band with
the Submillimeter Array. Although the angular resolution in their study is limited
to $\sim$$1^{\prime \prime}$, they suggested a value of RM near the central black
hole to be within $(-7.5~{\rm to}~3.4)\times 10^{5}\,{\rm rad\,m^{-2}}$, by
assuming that the bulk of the 230\,GHz emission originates in the jet base within
$\sim$$21\,R_{\rm s}$ from the black hole.

The detection of polarized signals in our 86\,GHz VLBI images provides some
important insights into the close vicinity of the central black hole of
M87. First, the detection of a polarized feature at $\sim$0.1\,mas
($\sim$14\,$R_{\rm s}$, projected) from the black hole provide evidence that the
RM associated with this feature is not larger than a certain value. If we assume
that the depolarization has an external origin and that the observed fractional
polarization degree is given by $m(\lambda) = m_0 \exp{(-2\sigma^2_{\rm
RM}\lambda^4)}$~\citep[where $m$, $m_0$ and $\sigma_{\rm RM}$ are observed
fractional polarization, intrinsic polarization and standard deviation of the RM
fluctuations, respectively;][]{burn1966}, the maximum $|\sigma_{\rm RM}|$ is
constrained to be $\sim$(5.8--17)$\times 10^4\,{\rm rad\,m^{-2}}$ (adopting
$\sim$$e^{-1}$--$e^{-3}$ damping of $m_{0}$). This is consistent with $|{\rm RM}|
< 7.5\times 10^5\,{\rm rad\,m^{-2}}$ obtained at a similar scale by
\citet{kuo2014}.  As a future work more dedicated VLBI polarimetric studies would
be fruitful to reveal the more detailed spatial distribution of RM near the jet
base.

The second point is that we detected a fractional polarization of $\sim$20\% in
the region at $\sim$0.4\,mas from the core, which is the highest value ever seen
in the pc-to-subpc scales of this jet. Since observed polarized signals are a
consequence of multiple depolarization effects under propagation
(internal/external Faraday depolarization, bandwidth/beam depolarization), the
intrinsic fractional polarization must be higher than the observed one (e.g., in a
manner mentioned above). Therefore, the observed highly polarized signal indicates
the presence of a well-ordered magnetic field in this region. The origin of such
an ordered field is still not clear, but a possible hint from our observation is
that this feature is located at the boundary of the southern limb (``sheath'') and
the central valley of emission (``spine''), where the presumed velocity shear may
amplify the longitudinal field
component~\citep[e.g.,][]{laing1980}. Interestingly, the observed EVPA for this
feature (that is perpendicular to the jet) seems to be consistent with the shear
interpretation, although the accuracy of the present EVPA calibration is still not
fully adequate to confirm this scenario. Alternatively, the ordered field could be
associated with a global helical/twisted field geometry, as predicted by the
magnetically-driven jet scenario~\citep[e.g.,][]{broderick2009}.

Ultimately, the further increase of the array sensitivity to total/polarized
intensity is necessary, as well as the implementation of a better polarimetric
calibration strategy. The upcoming incorporation of the Large Millimeter
Telescope~\citep{hughes2010} and of the phased-up ALMA to the millimeter VLBI
network is enormously powerful in this respect~\citep{fish2013}. This will allow
us to image the magnetic field structure in the jet-launching site as well as the
surrounding accretion flow structure in much more detail.

\section{Summary}

We reported results obtained from a new high-sensitivity, high-resolution VLBA+GBT
observation of the M87 jet at 86\,GHz. We summarize our main results as follows.

\begin{enumerate}
 \item We obtained a high-quality image of the jet-launch region of the M87 jet
       down to $\sim$10\,$R_{\rm s}$ near the black hole. The resulting image
       dynamic range is greater than 1500 to 1, which is the highest ever obtained
       for this source at 86\,GHz. The high-sensitivity image clearly confirmed
       some important well-known features of this jet such as a wide-opening angle
       jet launch, a limb-brightened intensity profile, a parabola-shape
       collimation profile and a counter jet. The limb-brightened structure is
       already well developed at $<28$\,$R_{\rm s}$ (projected) from the core, and
       the corresponding apparent full-opening angle near the black hole becomes
       as broad as $\sim$100$^{\circ}$. This is the broadest opening angle ever
       seen in any astrophysical jets as well as in the M87 jet itself.

 \item We discovered a complicated jet launch shape near the black hole
       ($r\lesssim$100\,$R_{\rm s}$) in our 86\,GHz image, indicating multiple
       collimation stages before the jet finally reaches the well-defined parabola
       profile in the larger scale. In particular, there is a ``constricted''
       structure at $\sim$35\,$R_{\rm s}$ (projected) from the core, where the jet
       cross section is locally shrinking. We suggest that an external pressure
       support/contribution from the inner part of accretion flow (presumably an
       ADAF type hot accretion flow or associated corona) may be dynamically
       significant in shaping and confining the M87 jet on this scale.

 \item Complementing our 86\,GHz data with close-in-time multi-epoch
       lower-frequency data, we detected proper motions in both the main jet and
       the counter jet, which were all subrelativistic. A mean speed of the main
       jet components were $\beta_{\rm app}$$\sim$0.32, while the counter jet was
       slightly slower at $\beta_{\rm app}$$\sim$0.17. Comparing the measured
       proper motions in the jet and the counter jet, the viewing angle for the
       inner jet is estimated to be $\theta$$\sim$29--45$^{\circ}$, although the
       more dedicated proper motion studies are necessary.

 \item We reported on the first VLBI 86\,GHz polarimetric result of the M87
       jet. While it is still challenging to reveal the entire polarimetric
       property of this jet, we detected some polarized features near the jet base
       at this frequency. The detection of the polarization signals at this
       frequency implies that the magnitude of the rotation measure toward these
       features are not larger than $\sim$(5--17)$\times10^4\,{\rm rad\,m^{-2}}$,
       which is consistent with the result reported in the recent 230\,GHz
       polarimetric study. Moreover, one of the polarized features has an observed
       fractional polarization up to $\sim$20\%, which is the highest value ever
       seen on pc-to-subpc scales of this jet. This indicates the presence of a
       well-ordered magnetic field in the formation and collimation zone of the
       M87 jet.
\end{enumerate}

\acknowledgments 

We sincerely thank the anonymous referee for his/her careful reviewing for
improving the manuscript. We also thank Jos\'e-Luis G\'omez for his valuable
comments on the polarimetric analysis; and Shin Mineshige for useful
discussion. The Very Long Baseline Array and the Green Bank Telescope are operated
by the National Radio Astronomy Observatory, a facility of the National Science
Foundation, operated under cooperative agreement by Associated Universities,
Inc. This work made use of the Swinburne University of Technology software
correlator~\citep{deller2011}, developed as part of the Australian Major National
Research Facilities Programme and operated under license. This study makes use of
43\,GHz VLBA data from the VLBA-BU Blazar Monitoring Program (VLBA-BU-BLAZAR;
http://www.bu.edu/blazars/VLBAproject.html), funded by NASA through the Fermi
Guest Investigator Program. This work was partially supported by KAKENHI
(26800109). Part of this work was done with the contribution of the Italian
Ministry of Foreign Affairs and University and Research for the collaboration
project between Italy and Japan. K.H. is supported by the Research Fellowship from
the Japan Society for the Promotion of Science (JSPS).

\appendix
\section{The calibrator 3C\,273}

Here we describe our data analysis and imaging for the bright quasar 3C\,273
obtained by the 86\,GHz VLBA+GBT program plus an additional 43\,GHz archival
dataset. Checking the total-intensity and polarimetric status of this source is
important for validating the results for M87 (particularly the polarimetric one).

\begin{figure}[htbp]
 \centering
 \includegraphics[angle=0,width=1.0\columnwidth]{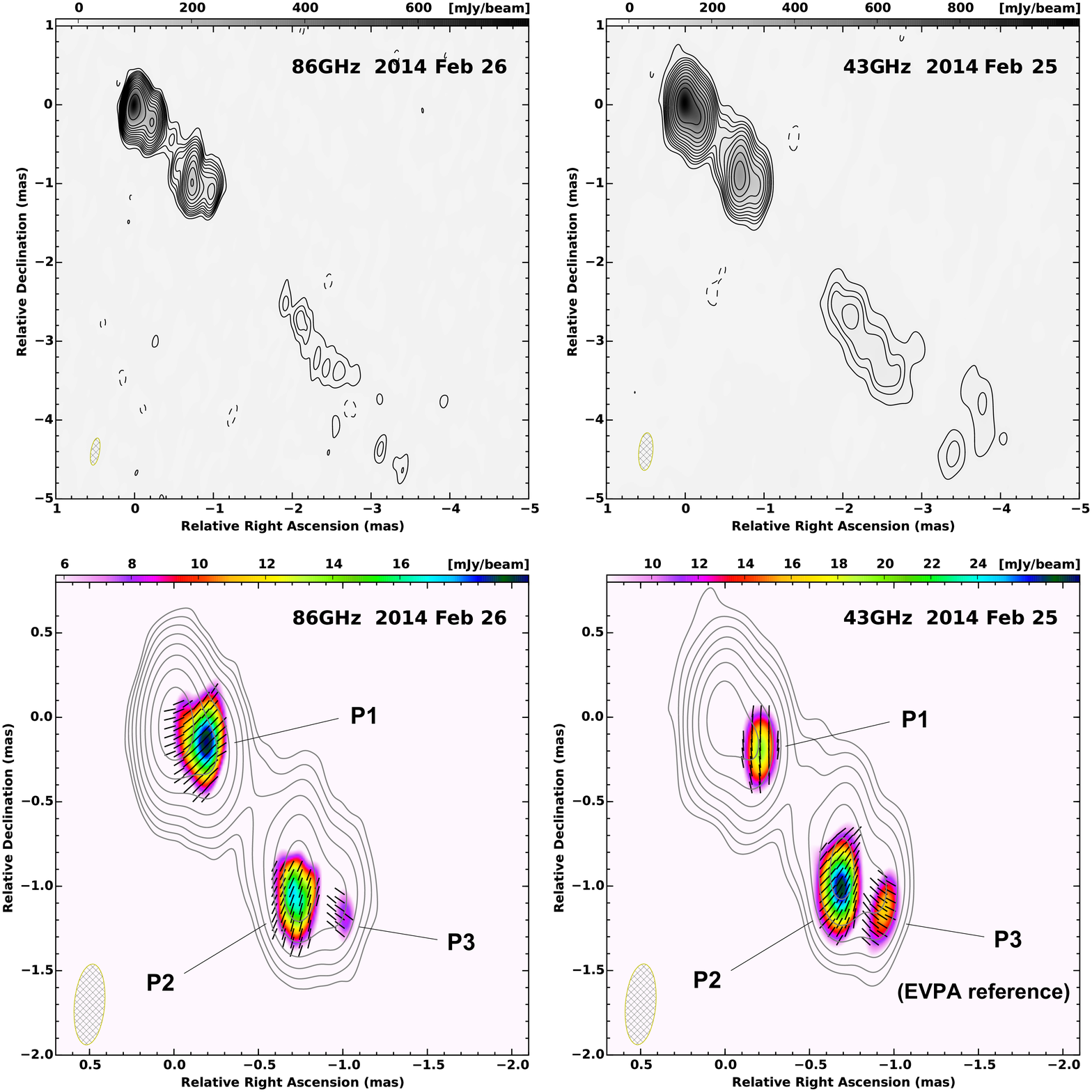}
 \caption{Total intensity and polarization images of 3C\,273 at 86 and
 43\,GHz. The top-left panel is a VLBA+GBT 86\,GHz total intensity contour image
 observed on 2014 February 26. The beam size and the peak intensity are $0.34
 \times 0.11$\,mas at PA $-10^{\circ}$ and 805\,${\rm mJy\,beam^{-1}}$,
 respectively. The top-right panel is a 43\,GHz VLBA total intensity contour image
 observed on 2014 February 25. The beam size and the peak intensity are $0.48
 \times 0.18$\,mas at PA $-5^{\circ}$ and 1100\,${\rm mJy\,beam^{-1}}$,
 respectively. The bottom-left and the bottom-right panels are a close-up view of
 the inner jet at 86 and 43\,GHz, respectively. In the bottom two panels, the
 images are restored with a common beam of $0.48\times 0.18$\,mas in PA
 $-5^{\circ}$, corresponding to the synthesized beam of the 43\,GHz data. The
 color map, the vectors and the contours indicate the observed polarized
 intensity, the observed EVPA and the total intensity distribution, respectively.
 P1, P2 and P3 indicate the polarized components which are consistently identified
 in both images. For the upper panels contours start from -1, 1, 2, 2$^{3/2}$, 4,
 2$^{5/2}$... times 3\,$\sigma$ rms level of each image ($1\,\sigma=1.2\,{\rm
 mJy\,beam^{-1}}$ and 3.3\,${\rm mJy\,beam^{-1}}$at 86 and 43\,GHz, respectively),
 while for the bottom panels contours are 1, 2, 4, 8... times 3.6\,${\rm
 mJy\,beam^{-1}}$ and 9.9\,${\rm mJy\,beam^{-1}}$ at 86 and 43\,GHz,
 respectively.} \label{fig:3c273}
\end{figure}

3C\,273 was observed on 2014 February 11 and 26 with VLBA+GBT at 86\,GHz as an
overall calibrator of our program. In the top-left panel of Figure~\ref{fig:3c273}
we show an 86\,GHz total intensity image of 3C\,273 taken on February 26. The
observed jet structure consists of the bright core with several discrete knots
down the jet. We achieved a dynamic range of 670 to 1, allowing a firm detection
of the weaker emission down to $\sim$5\,mas from the core. For comparison, in the
top-right panel of Figure~\ref{fig:3c273} we show a close-in-time VLBA 43\,GHz
image of this source that was observed on 2014 February 25 as a part of the Boston
University blazar monitoring program. The dynamic range of the 86\,GHz image is
about twice greater than that of the 43\,GHz one. This enables a comparison of the
images at a high confidence, despite the steep-spectral nature of the synchrotron
emission. The overall jet structure and the positions of the individual features
in these two images are in excellent agreement with each other.

Regarding polarimetry, our results are summarized in the bottom two panels of
Figure~\ref{fig:3c273}. The bottom-left panel shows the result obtained from the
VLBA+GBT 86\,GHz observation on 2014 February 26, while the bottom-right panel is
the result from the VLBA 43\,GHz observation on 2014 February 25. Within the
central 2\,mas of the jet, we identified three prominent polarized components that
were consistently detected in the 86 and 43\,GHz images. As described in
Figure~\ref{fig:3c273} these are termed as P1, P2 and P3 starting from the
upstream side. At 43\,GHz the highest polarized intensity is in P2, while P1 is
the strongest polarized component at 86\,GHz. At both frequencies the largest
fractional polarization was in P3 (10--15\%), while P1 was the smallest
(4--7\%). The observed positions of P2 and P3 are in good agreement in both 43 and
86\,GHz images. For the innermost component P1, the peak position of the polarized
flux is slightly offset between 43 and 86\,GHz, in the sense that the polarized
emission at 86\,GHz can be seen closer to the core than that at 43\,GHz. As
described below, this is presumably due to the decrease of the Faraday
depolarization effect at the higher frequency.

In terms of EVPA, we made use of the 43\,GHz polarization image as a reference for
our EVPA correction for the 86\,GHz data. Here we assume that the EVPA of the
outermost component P3 is stable with both time and frequency, and performed a
nominal EVPA correction by matching the observed 86\,GHz EVPA of P3 to that of the
43\,GHz one. The EVPA map shown in the bottom-left panel of Figure~\ref{fig:3c273}
was produced through this procedure. As a result, one can see that the EVPA of P2
is also closely aligned between the two frequencies. This indicates that the
absolute difference in rotation measure between P3 and P2 is relatively small.

On the other hand, the previous concurrent 43/86\,GHz polarimetric VLBA study of
this source suggested a large RM ($\sim$2$\times 10^4\,{\rm rad\,m^{-2}}$) for the
inner jet at $\sim$0.8\,mas from the core~\citep{attridge2005}. In fact, looking
into our innermost component P1, we can find a notable difference in EVPA
direction between the two frequencies, in contrast to the good EVPA alignment at
P2 and P3. The observed EVPA difference in P1 between 43/86\,GHz (in the
assumption that the P3's EVPAs are aligned between the two frequencies) is
$\Delta\chi = \chi_{\rm 7mm} - \chi_{\rm 3.5mm} = 50^{\circ}$. This indicates that
there is an absolute difference in RM between P3 and P1 of at least $|\rm{RM}| =
2.4 \times 10^4\,{\rm rad\,m^{-2}}$. This level of RM is very similar to that
reported by \citet{attridge2005}. It seems that there is an additional (i.e.,
another $\sim$15$^{\circ}$) rotation of EVPA at the near-core side of P1 in the
86\,GHz image. This implies the further increase of RM toward the core, which is
consistent with the non-detection of polarized signal in the corresponding region
at 43\,GHz.

To estimate the absolute uncertainty of our EVPA correction procedure, it is
necessary to know the absolute difference of P3's EVPA between 43 and
86\,GHz. Unfortunately this is difficult to derive from the present data alone. If
P3 has a similar level of RM as seen toward P1, the EVPA distribution shown in the
bottom-left panel of Figure~\ref{fig:3c273} will have another $\sim$$50^{\circ}$
rotation. However, this level of uncertainty should be regarded as an upper limit,
and the actual uncertainty should be smaller than this value, since P3 is located
further downstream and shows the higher fractional polarization than that of P1,
favoring the less amount of Faraday screen toward P3. On the larger scales (i.e.,
from a few to 10\,mas from the core), several authors reporte RM values of
hundreds to a few thousands rad\,m$^{-2}$~\citep[e.g.,][]{asada2002, zavala2005}.
Looking into the adjacent component P2, there is actually a slight
($\sim$$8^{\circ}$) difference in EVPA between 43 and 86\,GHz. This suggests an
absolute difference in $|\rm{RM}|$ between P3 and P2 of $3.9\times 10^3\,{\rm
rad\,m^{-2}}$, which is just in between the values for P1 and the outer jet in the
literature. Therefore, we regard that a level of $\sim$$4\times 10^3\,{\rm
rad\,m^{-2}}$ would be a reasonable measure of $|\rm{RM}|$ at P3, and thus a
likely uncertainty of our 86\,GHz EVPA correction relative to the 43\,GHz data
would be $\Delta \chi_{\rm RM}\sim 8^{\circ}$. In summary, adopting that the EVPA
uncertainty of the Boston 43\,GHz map is $\Delta \chi_{\rm 43GHz}\sim
10^{\circ}$~\citep{jorstad2005}, we estimate that a potential total uncertainty of
EVPA in our 3C\,273's 86\,GHz images is a level of $\Delta \chi_{\rm 3C273} \sim
\Delta \chi_{\rm 43GHz} + \Delta \chi_{\rm RM} \sim \pm 18^{\circ}$.

\end{document}